# Chemical trends of deep levels in van der Waals semiconductors

Penghong Ci[1,2], Xuezeng Tian[3], Jun Kang[4], Anthony Salazar[1], Kazutaka Eriguchi[1], Sarah Warkander[1], Kechao Tang[1], Jiaman Liu[1], Yabin Chen[1,5], Sefaattin Tongay[6], Wladek Walukiewicz[2], Jianwei Miao[3], Oscar Dubon[1,2], Junqiao Wu[1,2]*

[1]Department of Materials Science and Engineering, University of California, Berkeley, California 94720, United States.

[2]Materials Sciences Division, Lawrence Berkeley National Laboratory, Berkeley, California 94720, United States.

[3]Department of Physics & Astronomy and California NanoSystems Institute, University of California, Los Angeles, CA, USA.

[4]Beijing Computational Science Research Center, Beijing, China.

[5]School of Aerospace Engineering, Beijing Institute of Technology, Beijing, China.

[6]School for Engineering of Matter, Transport, and Energy, Arizona State University, Tempe, Arizona 85287, United States.

**Electronic mails to whom correspondence should be addressed:**

wuj@berkeley.edu

## Abstract

**Properties of semiconductors are largely defined by crystal imperfections including native defects. Van der Waals (vdW) semiconductors, a newly emerged class of materials, are no exception: defects exist even in the purest materials and strongly affect their electrical, optical, magnetic, catalytic and sensing properties. However, unlike conventional semiconductors where energy levels of defects are well documented, they are experimentally unknown in even the best studied vdW semiconductors, impeding the understanding and utilization of these materials. Here, we directly evaluate deep levels and their chemical trends in the bandgap of $MoS_2$, $WS_2$ and their alloys by transient spectroscopic study. One of the deep levels is found to follow the conduction band minimum of each host, attributed to the native sulfur vacancy. A switchable, DX center - like deep level has also been identified, whose energy lines up instead on a fixed level across different hosts, explaining a persistent photoconductivity above 400K.**

# Introduction

Defects with energies falling within the bandgap may act as a trap or emitter of free charge carriers[1], a site for exciton recombination[2], and a center to scatter electrons or phonons[3]. In conventional semiconductors, native defects such as vacancies introduce levels close to the middle of the bandgap when the material is more covalently bonded, or close to the band edges when the material is more ionically bonded, resulting in the former materials being defect sensitive while the latter materials are relatively defect tolerant[4]. Comparing positions of defect levels across different host materials helps to reveal chemical trends that inform defect models with broad impact. For example, the deep level associated with a given impurity[5] or native defect[6] tends to lie universally at a fixed energy position with respect to the vacuum level even when doped in different semiconductors, which can be used to determine band alignments of the host materials; equilibrium native defects tend to drive the Fermi level toward a stabilization position, and this position with respect to the bandgap can be used as a descriptor of doping propensity and doping limit of the semiconductor[7]; the DX center, an metastable defect switchable between deep and shallow states, dominates the free electron density in III-V semiconductor alloys[8]. It is critical to ask whether such insights and knowledge attained in studying conventional semiconductors are applicable in vdW materials. New effects of defects may emerge because the layered nature of vdW materials allows stronger lattice relaxation as well as new types of defects such as intercalated atoms.

Scanning tunneling microscopy (STM) is able to experimentally visualize various types of defects on the surface and relate these imperfections to electronic structures in vdW crystals[9], in particular for the most abundant native point defects that play a critical role in their electrical[10-13], optical[2], magnetic[14], catalytic[15] and sensing properties[16]. However, STM studies have led to inconsistency on the defect types with transmission electron microscopy investigations, as well as discrepancy in signatures of defect-induced mid-gap states from theoretical calculations[1,3,17-21], largely because of unclear differentiation of STM contrast between the metal and chalcogen sublattices and the complicated convolution of electronic and geometric structures[9]. Furthermore, it shows very limited capability in detecting defects beneath the surface.

In this work, we use deep level transient spectroscopy (DLTS), a high-frequency capacitance transient thermal scanning method[22,23], to characterize electronic structures of the deep traps inside the bandgap of vdW semiconductors, particularly $MoS_2$, $WS_2$ and their alloys, including their energy positions and capture cross sections. Combined with atomic-resolution scanning transmission electron microscopy and first-principles calculations, one of the deep levels determined by DLTS is identified as sulfur vacancies, whose energy position follows the conduction band edge in the host materials, distinct from vacancy defects in traditional group III-V semiconductors. A metastable DX center is identified in these vdW semiconductors, featuring a persistent photoconductivity above 400 K and explaining the chemical trend of native electron concentration in the hosts.

## Results

**DLTS devices and DLTS spectra.**

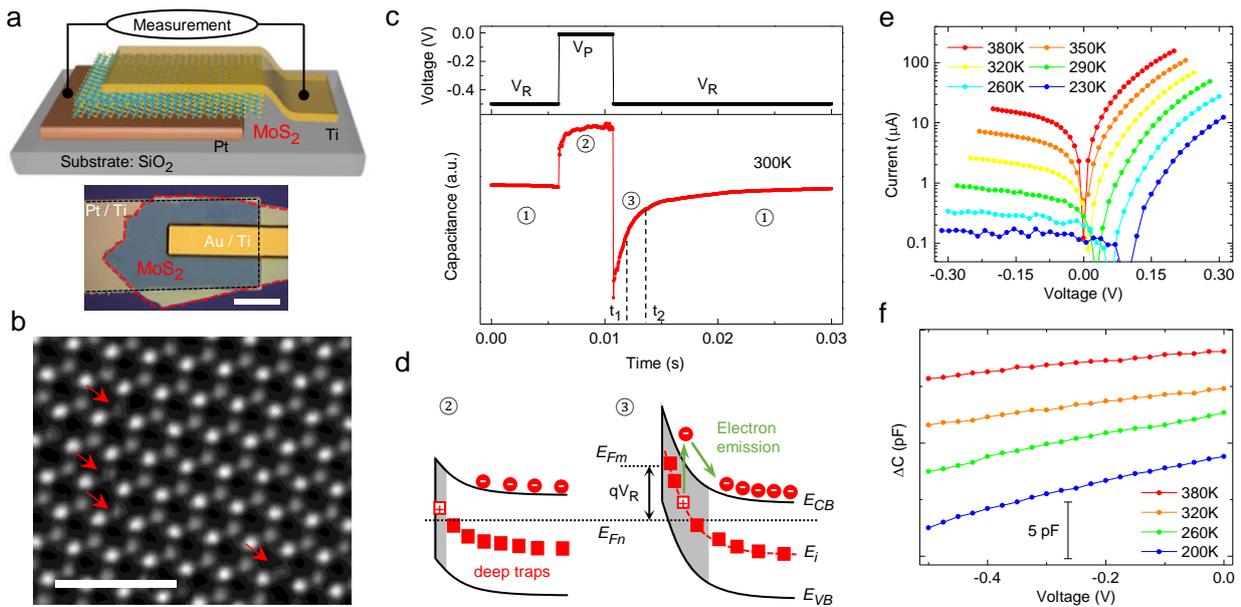

**Figure 1: Materials and devices for transient spectroscopic study of defects. a**, Schematic and optical

image (scale bar: 20um) of an asymmetric $MoS_2$ device for DLTS, with Schottky contact ($MoS_2$/Pt/Ti) on the bottom and Ohmic contact (Au/Ti/$MoS_2$) on the top. **b**, Aberration-corrected STEM image of a monolayer $MoS_2$ exfoliated from the materials used for devices. Red arrows highlight S vacancies ($V_S$). Scale bar, 1nm. **c**, Capacitance transient (bottom) in response to a pulsed change in bias voltage (top). **d**, Band bending of the Schottky junction ($MoS_2$/Pt), illustrating the electron trapping (②) and emission process (③) of deep traps in the depletion region (shaded). $V_R$ tunes the Fermi level of the n-type $MoS_2$ ($E_{Fn}$) with respect to that of the metal contact ($E_{Fm}$). **e** & **f**, Temperature-dependent I-V and C-V curves confirming the Schottky-Ohmic contacts.

Mechanically exfoliated, multilayer (~ 50 nm) flakes of freshly grown $Mo_{1-x}W_xS_2$ ($x$ = 0, 0.4, 0.7, 1) crystals were made into two-terminal Schottky-Ohmic devices (Fig. 1a). The Schottky contact was formed by dry-stamping freshly exfoliated flakes onto pre-deposited Pt electrodes and confirmed by the *I-V* and *C-V* curves shown in Fig. 1e and 1f, both of which show the n-type conductivity of $MoS_2$. This maximally protects the depletion region at the Schottky contact against contamination and damage[24], as it is at this region where the deep levels trap and emit charge carriers during the DLTS measurement. The measured total capacitance (Supplementary Fig. 10) is composed of that of the DLTS device ($C_{device}$) and the stray capacitance ($C_{stray}$) connected in parallel. The latter, although with a large value, is insensitive to the external differential voltage (Supplementary Fig. 10), hence the variation of capacitance under the biased voltage indeed probes the former (Fig. 1f).

The depletion width at the Schottky junction (~ 20 nm, the shadow in Fig. 1d), hence the capacitance (Fig. 1c), is initially held constant by a steady-state reverse bias ($V_R$ = - 0.5 V, stage ①)[23]. An opposite voltage pulse ($V_P$) is then added onto $V_R$, reducing the depletion width (as evidenced by the increased capacitance at less-negative voltage, Fig.1f), and allowing the traps in the initial depletion region to be filled with free electrons (stage ②)[23]. When the initial, constant bias is restored, the return of the capacitance to the steady-state value is characterized by a transient (stage ③) related to the emission of majority carriers from the deep traps in the material. The capacitance difference within a rate window (between the pre-set $t_2$ and $t_1$ in Fig. 1c)[22] reaches the maximum at a specific temperature. The emission rate ($e_n$) in stage ③ depends exponentially on temperature via the trapping energy level ($E_i$) measured from the conduction band minimum (CBM, $E_{CB}$)[22],

$$\frac{e_n}{T^2} = K\sigma_n \exp\left(-\frac{|E_{CB}-E_i|}{k_B T}\right) \quad (1)$$

where $\sigma_n$ is the capture cross section, and $K$ is a known constant. Arrhenius plots of Eq.(1) at various rate windows (0.5 ms to 20 ms in Fig. 2a) allow extraction of the activation energy of deep levels, $E_{CB}$ - $E_i$. For MoS$_2$ we found two, 0.27 ± 0.03 eV (peak A) and 0.40 ± 0.02 eV (peak B), as shown in Fig. 2b. The positively valued DLTS peaks (Fig. 2a) indicates that these are majority carriers traps in MoS$_2$[22]. We also measured current transient spectroscopy (CTS, see Supplementary Fig. 1) by recording the current rather than capacitance under the pulsed bias[25], yielding an activation energy of $E_{CB}$ - $E_i$ = 0.25 ± 0.02 eV for MoS$_2$ (Supplementary Fig. 1a), consistent with the peak A in DLTS. We note that for each of the trap energies obtained in this work, at least two devices were measured and all show consistently very similar energy. Thermodynamically, the slope of Eq. (1) corresponds to the change of enthalpy (ΔH), different from the Gibbs free energy ΔG ( = $E_{CB} - E_i$)[26], but the difference can be neglected when electrons are excited from the traps to the conduction band without invoking changes in the bonding configuration (see Supplementary Note 5)[27].

**Determination of sulfur vacancies from STEM and DFT calculations.**

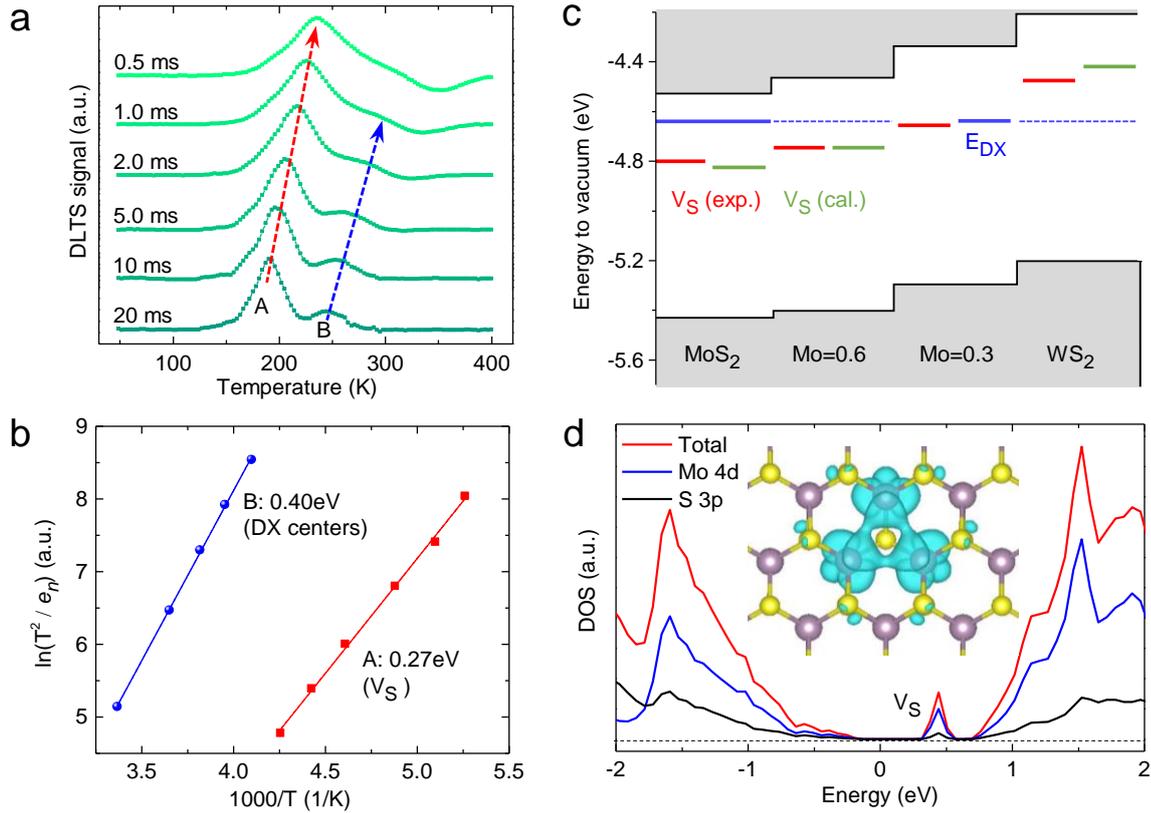

**Figure 2: Deep levels and their alignment in vdW crystals. a,** DLTS signal of a MoS$_2$ device at different rate windows, and **b,** the resultant Arrhenius plots to determine the activation energies. **c,** Conduction and valence band edge alignment calculated with respect to the vacuum level, and positions of deep level experimentally identified in this work. Solid black lines: CBM and VBM in bulk crystals from our DFT calculations; red bars: deep levels attributed to V$_S$ measured by DLTS and CTS; blue bars: DX center levels determined by DLTS and PPC (dashed blue line is guide to the eye); green bars: DFT-calculated V$_S$ levels. **d,** density of states (DOS) for multilayer MoS$_2$ with V$_S$. Inset: real space distribution of the wavefunction of V$_S$ state. The purple and yellow spheres represent Mo and S atoms, respectively.

To reveal the atomic origin of these deep traps, we have performed first-principles calculations of S single vacancies (V$_S$) in multilayer MoS$_2$, WS$_2$ and their alloys. V$_S$ is chosen because it is the most abundant defect known to naturally occur in these materials[21]. The calculation shows that V$_S$ would introduce a deep-level state with energy of 0.29 eV (for MoS$_2$) and 0.21 eV (for WS$_2$) below the CBM, in good agreement with the DLTS/CTS results. We note that the value of 0.29 eV is also consistent with the calculated $V_S$ energy in $MoS_2$ previously reported in literature[1,3]. Our calculations also confirm that $V_S$ is a deep acceptor, labeled as (0/-)[1,23], not responsible for the natively n-type conductivity of $MoS_2$. The neutral ground state implies its extremely weak Coulomb attraction to electrons, and hence very small capture cross section. $V_S$ defects are directly observed in these materials by scanning transmission electron microscopy (STEM, Fig. 1b and Supplementary Fig. 5)[11], where the density of $V_S$ is directly determined to be $1\sim3 \times 10^{20}$ cm$^{-3}$ (Supplementary Fig. 4), on the same order of those reported in literature[3,21]. The STEM study also confirms that $V_S$ is the dominant point defects, and no other defects or impurities were detected in the materials. We note that akin to conventional semiconductors, not all of these $V_S$ are electronically active (Supplementary Fig. 6); in fact, deep traps can be highly passivated or compensated, as observed in GaN and GaAs[28,29].

To reveal the chemical trend of the $V_S$ level in different vdW semiconductors, $Mo_{0.6}W_{0.4}S_2$, $Mo_{0.3}W_{0.7}S_2$, and $WS_2$ were also synthesized and then assembled into Schottky devices for similar DLTS/CTS measurements (Supplementary Fig. 1). All of these materials exhibit at least one deep level, akin to the feature A observed in $MoS_2$, with an energy level below the CBM of the host material of $0.29 \pm 0.02$ eV, $0.31 \pm 0.02$ eV and $0.26 \pm 0.04$ eV (red bars in Fig. 2c), respectively. These energy levels are all in good agreement with the DFT calculated $V_S$ levels, as shown by the green bars in Fig. 2c and the refined band structure with $V_S$ in Supplementary Fig. 8.

Some deep levels in different isovalent materials line up at a fixed position with respect to the vacuum level, such as oxygen dopant or Ga dangling bond in different $GaAs_{1-x}P_x$ alloys[6,30]. In contrast, the red bars in Fig. 2c show that as the W fraction increases in $Mo_{1-x}W_xS_2$, the energy level of $V_S$ shifts monotonically toward the vacuum level; that is, the $V_S$ level largely follows the CBM of the host. This is understandable because, as shown in the partial density of states plot in Fig. 2d, the $V_S$ state originates mostly from the 4d (5d) orbitals of the Mo (W) atoms, rather than the S atoms, sharing the same orbital composition as the CBM[31,32]. Following this finding, anion impurities (such as oxygen) substituting S are predicted to create deep levels also about 0.3 eV below the CBM of the host (see Supplementary Fig. 7), because it is known that highly electronegative, substitutional dopants tend to have similar wavefunctions as those of ideal vacancies[6]. The electron capture cross section ($\sigma_n$) of $V_S$ is evaluated from Eq. (1) to be ~$3.6\times10^{-18}$ cm$^2$ in MoS$_2$, using the thermal velocity effective mass (0.57 m$_o$) and effective density of states mass (0.50 m$_o$) obtained from our DFT calculation and literature (see Supplementary Note 1). This value is small but comparable to that of Zn acceptor level in Si and Cu acceptor level in Ge[23,33].

**Persistent photoconductivity and DX center model.**

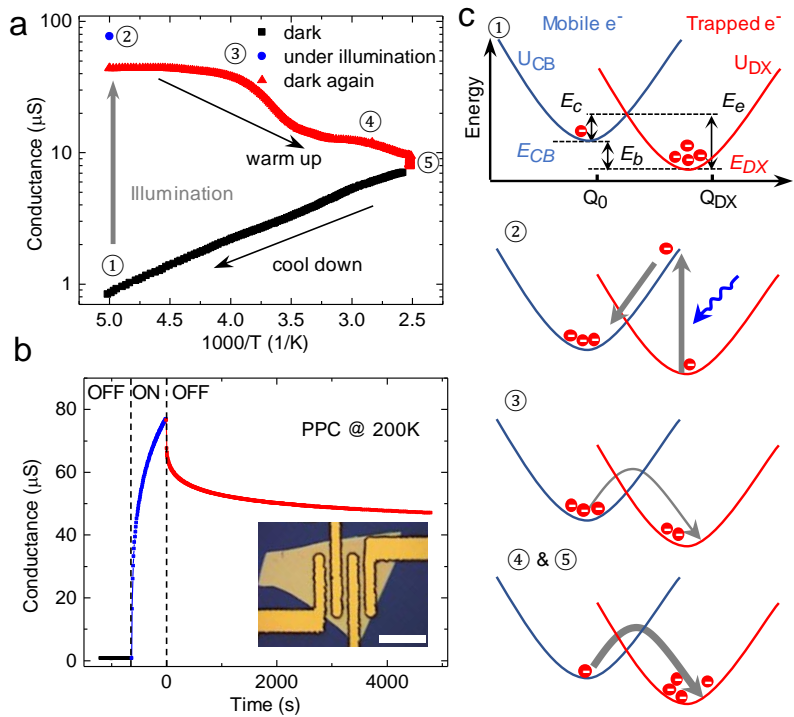

**Figure 3: Temperature-dependent persistent photoconductivity (PPC) and the DX center model. a**,

Conductance of a MoS$_2$ device as a function of temperature before, during, and after exposure to light illumination. **b**, PPC transient of the device at 200K. Inset: optical image of a four-probe device for PPC measurement (scale bar: 20μm). **c**, Configurational coordinate diagram (CCD) showing the three energies to characterize the DX center and describe the five PPC processes in **a**.

To explore the origin of peak B (0.40 eV) in MoS$_2$ shown in Fig. 2a, we obtained complementary information about deep levels from photoconductivity measurements. Photoconductivity, especially when it is persistent (persistent photoconductivity, PPC), has been used to gauge conduction by charge carriers photo-liberated from certain deep traps[34,35]. Figure 3a shows temperature-dependent dark conductance of a MoS$_2$ flake (~ 50 nm thick) measured in four-probe geometry (Inset of Fig. 3b). The sample was cooled in darkness from 400 K to 200 K (black data points). It was then exposed to white light for 10 minutes (blue data point) at 200 K, during which the conductance became two orders of magnitude higher than in the dark. When the light was switched off ("dark again") at this low temperature, the conductance dropped slightly, but still stayed > 50 times higher than the pristine dark state. The PPC stayed at this level for at least 11 hours at 200K (Fig.3b). When the sample was warmed up, the conductance stayed at the higher level (red data points) until a temperature of 400 K where it nearly converged to the pristine dark conductance.

Such a PPC effect in response to light exposure and temperature is a direct manifestation of metastability of defect states, and a hallmark of DX centers in semiconductors[34,35]. DX centers, observed in the 1980s in many III-V semiconductors such as AlGaAs, are a special type of localized states resonant with the conduction band of the host[8]. In contrast to ordinary deep levels, DX centers are capable of switching into a charge-delocalized, electron-donating state via significant lattice relaxation when triggered by external stimuli, such as light and gate control[8,36,37]. Typically described in the configurational coordinate diagram (CCD) as shown in Fig.3c, DX centers are characterized by a parabolic coordinate (Q) dependence of DX center energy ($U_{DX}$) intersecting that of the delocalized state ($U_{CB}$)[8]. The displacement along the Q axis between the two minima describes a large lattice relaxation that reflects the metastability of the DX centers. Three energies are thus defined: capture activation energy ($E_c$), which is the energy barrier for the DX center to trap an electron and can be determined from the kinetics of PPC; emission activation energy ($E_e$), the energy barrier to de-trap (emit) an electron, measured via DLTS[8,36]; and energy depth ($E_b = E_{CB} - E_{DX} = E_e - E_c$), which is the ground state energy ($E_{DX}$) measured from the CBM ($E_{CB}$) and can be derived from the temperature dependence of conductance.

As shown in Fig. 3a&c, at the thermal equilibrium state (stage ①), most electrons are trapped in the DX centers. Upon excitation by light with energies above the optical threshold (stage ②)[8,36], electrons in $E_{DX}$ are photo-excited to $E_{CB}$. When the light is off, these electrons stay in $E_{CB}$ and are blocked by the barrier $E_c$ from relaxing back to $E_{DX}$, causing the PPC (stage ③). When temperature rises, more electrons are thermally excited over $E_c$ into $E_{DX}$ (stage ④), eventually recovering to the pristine, dark-state conductivity (stage ⑤). In this study, the PPC effect exists at temperatures up to more than 400 K (upper limit of our equipment). This is in stark contrast to the PPC effect of DX centers discovered in group III-V semiconductors, where it survives only at $T < \sim 140$ K[35,37,38].

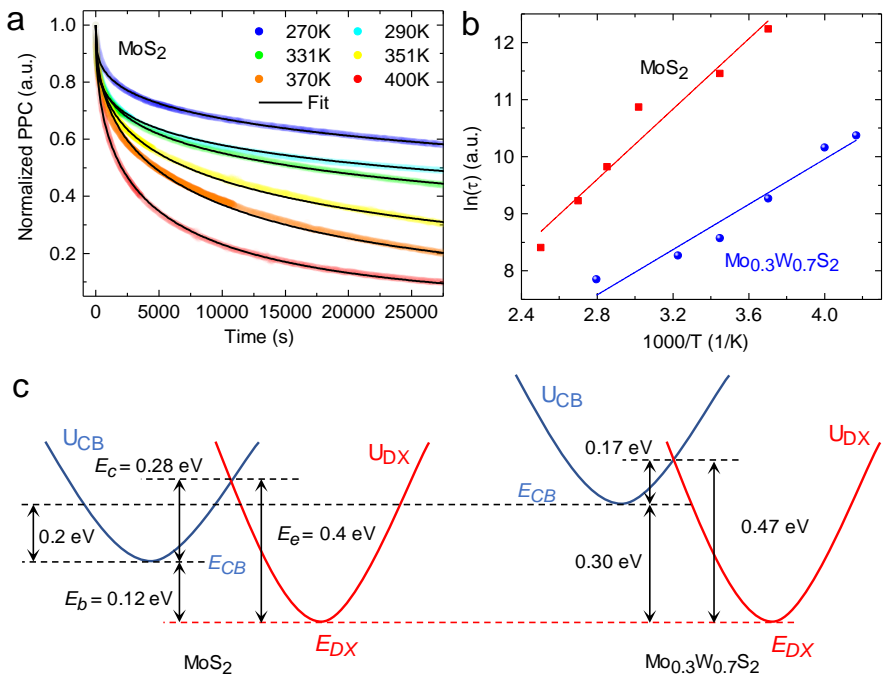

**Figure 4: DX center levels measured in MoS$_2$ and Mo$_{0.3}$W$_{0.7}$S$_2$. a**, Transient normalized-PPC curves at various temperatures for MoS$_2$ and **b**, the resultant Arrhenius plots of the time constant for MoS$_2$ and Mo$_{0.3}$W$_{0.7}$S$_2$. Semitransparent points: experimental data; thin solid lines: fitting to Eq. (2). **c**, CCD for MoS$_2$ and Mo$_{0.3}$W$_{0.7}$S$_2$, where the energy depth ($E_b$), capture ($E_c$) and emission ($E_e$) energy barriers are determined by temperature dependence of conductance, PPC and DLTS, respectively. The band offset between the two materials is obtained from DFT, resulting in a flat lineup of the DX center energy level ($E_{DX}$) across different host materials. The error range for these energies is estimated to be ± 0.04 eV.

The transient PPC curves are plotted in Fig.4a for a range of temperatures, where non-persistent photocurrent was excluded, dark current was subtracted and the remaining part was normalized by the value at $t = 0$, the moment the illumination is terminated. Note that in order to reset the initial dark current before taking each of these PPC curves, the samples were kept at 400 K for at least one day in a high vacuum (~10$^{-6}$ torr) to drain the extra electrons in $E_{CB}$. We see that, consistent with the DX center model (Fig. 3), high temperature expedites the kinetics of the PPC decay. Following the treatment in literature, the PPC can be well described by the stretched-exponential equation[34,35]:

$$I_{PPC}(t)/I_{PPC}(0) = \exp[-(t/\tau)^\beta] \qquad (2)$$

where $\tau$ is the characteristic decay time constant, $\beta$ is a decay index with a value between 0 and 1. Because of the underlying thermal activation process, the temperature dependence of $\tau$ is related to the trap barrier via $\tau \propto \exp(E_c/k_B T)$[34,35]. Arrhenius plots of the temperature-dependent $\tau$ yield $E_c$ of 0.28±0.02 eV for MoS$_2$ and 0.17±0.02 eV for Mo$_{0.3}$W$_{0.7}$S$_2$ (Fig. 4b and Supplementary Fig. 3). These values are higher than $E_c$ (~0.14 eV) of DX centers reported in the Se-doped AlGaAs system[8], presumably because the layered structure of the vdW materials allows larger lattice relaxation than the tetrahedral structure of AlGaAs. The higher $E_c$ is also responsible for the extension of PPC to much higher temperatures.

The energy $E_b$ (= $E_{CB}$ - $E_{DX}$) characterizes the thermodynamic energy depth of the DX center, and was extracted from Arrhenius plots of the dark conductance of the sample (Supplementary Fig. 3). Values of $E_b$ = 0.12 eV and 0.30 eV were found for $MoS_2$ and $Mo_{0.3}W_{0.7}S_2$, respectively. Adding $E_b$ to $E_c$ gives $E_e$, the emission barrier, of 0.39 eV and 0.47 eV for $MoS_2$ and $Mo_{0.3}W_{0.7}S_2$, respectively. These values are in very good agreement with the energies of peak B measured in DLTS for $MoS_2$ (0.40 ± 0.02 eV) and $Mo_{0.3}W_{0.7}S_2$ (0.47 ± 0.02 eV). Therefore, we attribute the peak B measured in DLTS to emission of electrons from the DX centers. We note that, unlike regular deep levels (such as the $V_S$ state) which have no capture/emission barriers, for DX centers, the Arrhenius plot of the DLTS spectrum extracts the emission barrier $E_e$ (Fig. 2c and 3c), rather than $E_b$ which is the separation of $E_{DX}$ directly measured from the conduction or valence band edges (see more in Supplementary Note 6)[23]. Following the CBM offset of ~ 0.3 eV between $MoS_2$ and $WS_2$ from our DFT calculation, the CBM ($E_{CB}$) of $Mo_{0.3}W_{0.7}S_2$ is interpolated to be higher than that of $MoS_2$ by 0.2 eV. Combining all these energy values, the energy of $E_{DX}$ shows an interestingly flat alignment across these two compositions, as plotted in the CCD in Fig.4c. It is not surprising to see that the $E_{DX}$ position is independent of the material composition because it is also constant for DX centers in AlGaAs across different alloy compositions[8,36,38]: in AlGaAs alloys, $E_{DX}$ is located universally at 3.8 eV below the vacuum level, and does not follow the CBM of the host material (in contrast to shallow defect levels). DX centers act as deep traps that result in different shallow donor doping efficiency in AlGaAs with different compositions[8]; similarly, the chemical trend of energy level of DX centers in the vdW semiconductors can explain the well-known, orders of magnitude higher native free electron density in undoped $MoS_2$ than in $WS_2$, as the DX centers are shallower in the former (details in Supplementary Fig. 9). When they are doped, these deep defects also largely determine the doping efficiency and dopability of these materials, as they can compensate the shallow dopants.

# Discussion

Although our multipronged experiments show clear evidence of DX centers in these vdW semiconductors, elucidation of the atomic origin of the DX centers requires further exploration including extensive first-principles calculations. However, the flat alignment of $E_{DX}$ provides a clue. In AlGaAs, the electron wavefunction of the DX center is extremely localized on an Al/Ga site surrounded only by and bonded only to the nearest As atoms; therefore, $E_{DX}$ is very insensitive to the change of Al fraction in the alloy[37,39]. Similarly, in $Mo_{1-x}W_xS_2$ alloys where $E_{DX}$ is independent of the cation composition $x$, it is likely that the DX centers neighbor only S atoms, hence are either impurity atoms substituting the cation, or small interstitial atoms bonded to S. For example, a potential candidate would be a defect complex involving hydrogen bonded to S, a dopant inevitably and unintentionally introduced during the growth. Indeed, hydrogen has been proposed to be a possible origin of n-type native conductivity in $MoS_2$ due to the formation of shallow levels[40].

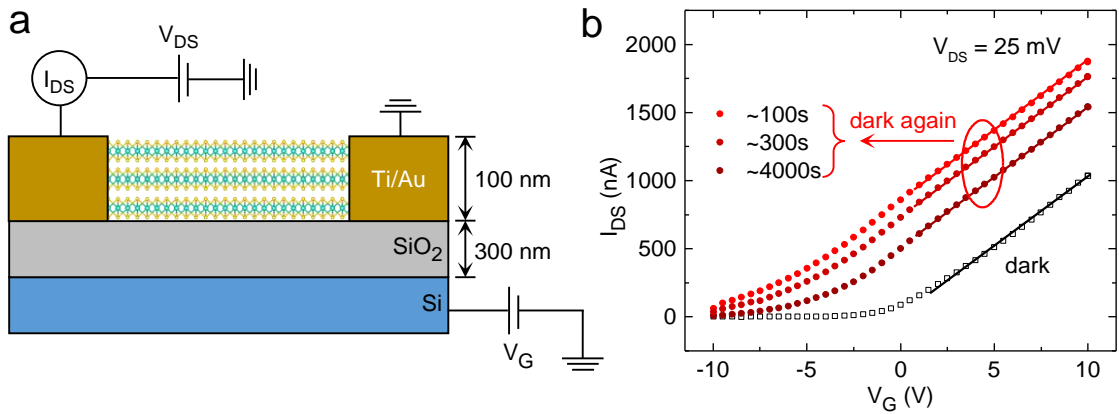

**Figure 5: Mobility of MoS$_2$ before and after illumination. a**, Schematic of a multilayer MoS$_2$ field-effect transistor (FET). **b**, Transient transfer characteristics for the FET before the illumination (hollow points) and at specific time after the illumination is turned off (solid points). The back gate voltage, $V_G$, is applied to the substrate. The solid lines show the slopes of the $I_{DS}$ - $V_G$ curves, corresponding to electron mobility of the channel material in the device.

The decay time constant describes the time it takes for the mobile electrons to be re-trapped by the DX centers, and dictates the relaxation kinetics of the free charge carrier density. The transient conductivity or current in the PPC (Fig. 4a) is assumed to have a similar relaxation kinetics as that of free carrier density, and hence can be used to extract the delay time constant for DX centers. This assumption is typically made in investigation of DX centers in traditional III-V semiconductors as the carrier mobility varies much less than the carrier density and the current is then directly proportional to carrier density [34-36,41,42]. In order to test the validity of this assumption in our case, a multilayer $MoS_2$ FET was made to determine the evolution of mobility before and after the illumination. Figure 5a depicts a multilayer $MoS_2$ assembled into the FET which is subjected to a back gate voltage ($V_G$). According to the data in Fig. 5b, we extract the low-field field-effect mobility to be ~ 16 cm$^2$/(V·S) for the $MoS_2$ channel, based on the expression $\mu = [dI_{DS}/dV_{Gate}] \times [L/WCV_{DS}]$ [43], where $W$ and $L$ are the width and length of the channel, respectively, $C$ is the capacitance of the gate dielectric $SiO_2$ layer, and $V_{DS}$ is the source-drain voltage in the FET. These $I_{DS}$ - $V_G$ curves exhibit the same slope before and after light illumination, indicating a constant mobility regardless of the density of free or trapped electrons in the channel, hence validating the extraction of time constant from electrical current via Eq. (2).

It is technically challenging to apply the DLTS to monolayers of vdW semiconductors, owing to expected high leakage current and issues arising from the sub-depletion width thickness. However, the deep levels we quantified for thick layers are expected to be applicable to monolayers and few layers. This is because the very weak interlayer vdW coupling only modulates the conduction and valence band edges, transitioning the material from direct bandgap in monolayers to indirect bandgap in the bulk, while hardly affecting the entire band structure[32]; on the other hand, the spatially very localized wavefunctions of deep levels do not only hybridize with the conduction or valence band edges, which is in contrast to shallow defects whose wavefunctions are composed of entirely the band edge states. For example, regarding monolayer $MoS_2$, first-principles calculations predicted that the $V_S$ deep level lies about 0.5 eV below the CBM at the *K* point in the Brillouin Zone[1,3,17,21,44], which is in good agreement with the 0.27 eV below the CBM at the *Q* point in multilayer $MoS_2$ quantified in this study, considering the 0.2 eV CBM offset between monolayer and bulk $MoS_2$ [45,46].

Our work determines energy levels and chemical trends of the most abundant native defects in $MoS_2$, $WS_2$ and their alloys. These energy levels offer quantitative references for both applications that are limited by defects such as transistors[10,24] and light emitting devices[2], as well as applications that are facilitated by defects such as catalysis[15] and sensors[16]. We also discover metastable and switchable, DX center-like defects in these vdW materials at temperatures above 400 K, in contrast to those in other semiconductors that exist only at T < 140 K[8,38]. As a result, practical device applications may be developed from the DX centers in vdW materials, such as nonvolatile memory based on a single defect. These defects may provide a platform for study of electron-phonon coupling, electron correlation, and many-body physics such as negative-U effects in quasi-two-dimensional crystals[30].

# Methods

**Materials preparation.** The vdW bulk crystals were synthesized using the flux zone technique without using transporting agent precursor, in order to reduce contamination[47]. The growth starts with 6N-purity, commercially available 300 mesh amorphous powders of molybdenum and/or tungsten and pieces of sulfur. Further electrolytic purification was necessary to eliminate magnetic impurities commonly found in metal powders, and secondary ion mass spectroscopy (SIMS) was used to test the purity. Powders were mixed at stoichiometric ratios, sealed under $10^{-7}$ torr pressure in quartz ampoules, and annealed up to 800°C for 10 days. The polycrystalline products were collected and resealed again. In the second formation process, a small temperature differential (~15°C) was created at high temperatures to thermodynamically drive the reactions. The crystallization process was slow and the entire growth was completed in a three-month time frame.

**Device fabrication.** Multilayer (~ 50 nm thick) $MoS_2$, $WS_2$ and their alloys were mechanically exfoliated from bulk crystals. For DLTS/CTS experiments, these samples were transferred onto Pt/Ti (45/10 nm) bottom electrodes[10], followed by photolithography, and electron beam evaporation of 20-nm Ti and then 80-nm Au as the top electrodes, and lift-off. In this way, the vdW flake is sandwiched by Pt (Schottky) metal at the bottom, and Ti (Ohmic) metal at the top. For PPC measurements, four-probe metal leads (Au (80nm) / Ti (20 nm), Ti at bottom) were deposited onto exfoliated samples. The devices used $SiO_2$ (300nm) / Si as the substrate.

**Electrical Measurements.** A deep level transient spectrometer (Sula Technologies) was used to measure DLTS, CTS, CV, and IV curves in Fig. 1 and 2. In this instrument, the emission rate is set as $e_n = 1/(D \times \Delta t)$, where $\Delta t = t_2 - t_1$ is the preset time difference in Fig. 1c and 2a, and $D$ is a constant representing the delay factor, 1.94 and 4.3 for the DLTS and CTS measurements, respectively. In the capacitance test, including CV and DLTS, an A.C. voltage with an amplitude of ~ 60 mV and frequency of 1MHz was superimposed onto the D.C. reverse bias. For the PPC measurements, four-terminal transport characteristics were measured by applying a DC bias to the outer channel and recording the current using a current amplifier and the voltage drop across the inner channel using a voltage amplifier. Optical illumination for the PPC was by a convection-cooled 30-Watt illuminator (Fiber-Lite 190).

**STEM characterization.** Mechanically-exfoliated monolayer $MoS_2$ was transferred from $SiO_2$ surface to TEM grids (Quantifoil R2/2) by selective etching of the $SiO_2$ in 49% hydrofluoric acid. Images were acquired from different regions of the monolayer $MoS_2$ using a Nion UltraSTEM 100 aberration-corrected STEM in ADF-STEM mode with E = 70 kV. The beam convergence semi-angle was 30 mrad and the detector collection angle was in the range of 30-300 mrad, where a small detector inner angle was chosen to reduce the electron dose. The energy spread of the electron beam was 0.3 eV. To reduce the total electron dose, images were measured with a beam current of 15 pA and a dwell time of 84 μs per image, which correspond to a total electron dose $4.7 \times 10^5$ e$^-$/Å$^2$. The ADF-STEM images contain a mixture of Poisson and Gaussian noise and were denoised by the block-matching and 3D filtering (BM3D) algorithm[48], from which S vacancies were identified. It has been reported previously that a 80 keV electron beam induces S vacancies in $MoS_2$ with a rate of $3.45 \times 10^8$ - $3.36 \times 10^9$ electrons per S vacancy[3,21]. As we used a 70 keV electron beam, the vacancy formation rate in our experiment should be $> 3.45 \times 10^8$ electrons per S vacancy. From the total electron dose used in our experiment, we estimated the electron beam induced S vacancy density in our sample was $< 2 \times 10^{20}$ cm$^{-3}$. Since we observed a S vacancy density of $3 \times 10^{20}$ cm$^{-3}$ in the $MoS_2$ sample, we concluded that the native S vacancy density was $> 1 \times 10^{20}$ cm$^{-3}$, which is in agreement with that of exfoliated undoped MoS$_2$ samples[21].

**DFT calculations.** The calculations were performed using the Vienna ab initio simulation package (VASP) with the projector-augmented wave method[49,50]. The generalized gradient approximation of Perdew-Burke-Ernzerhof (GGA-PBE) was adopted for the exchange-correlation functionals[51]. The energy cutoff for the plane-wave expansion was set to 350 eV. Structure relaxation was stopped when the force on each atom was smaller than 0.01 eV/Å. The van der Waals interaction was included by using the correction scheme of Grimme[52].

For defect calculations in bulk MX2, we employed 5×5×1 supercell, where a tilted c lattice vector was adopted, with c=c$_0$ + 2a$_0$ + 2b$_0$, where a$_0$, b$_0$, and c$_0$ are the primitive cell lattice vectors. As discussed in previous studies[1], this improves the convergence of total energies with respect to cell size. The k-point sampling is 2×2×2. The defect charge-transition energy level $\epsilon(q/q')$ corresponds to the Fermi energy $E_F$ at which the formation energy for a defect α with different charge state q and q′ equals with each other. It can be calculated by[53]:

$\epsilon(q/q') = [E(\alpha,q) - E(\alpha,q') + (q-q')(E_{VB} + \Delta V)]/(q'-q)$.

Here E(α,q) is the total energy of the supercell containing the defect, and $E_{VB}$ is the valence band maximum (VBM) energy of the host material. The potential alignment correction term $\Delta V$ is added to align the VBM energy in systems with different charged states. It is calculated by the energy shift of the 1s core-level energy of a specified atom (which is far away from the defect site) between the neutral defect and charged cases. For Mo$_{1-x}$W$_x$S$_2$ alloys, different S vacancy sites have different local environments. The number of surrounding Mo and W atoms varies, resulting in four types of V$_S$. We calculated the charge-transition levels for each type, and then carried out an average according to the concentration of different types to obtain the final charge-transition level.

**Data availability**

The data that support the plots in this paper are available from the corresponding author upon reasonable request.

References

1   Komsa, H.-P. & Krasheninnikov, A. V. Native defects in bulk and monolayer MoS 2 from first principles. *Physical Review B* **91**, 125304 (2015).
2   Amani, M. *et al.* Near-unity photoluminescence quantum yield in MoS2. *Science* **350**, 1065-1068 (2015).
3   Qiu, H. *et al.* Hopping transport through defect-induced localized states in molybdenum disulphide. *Nature communications* **4**, 1-6 (2013).
4   Pandey, M. *et al.* Defect-tolerant monolayer transition metal dichalcogenides. *Nano letters* **16**, 2234-2239 (2016).
5   Langer, J. M. & Heinrich, H. Deep-level impurities: A possible guide to prediction of band-edge discontinuities in semiconductor heterojunctions. *Physical review letters* **55**, 1414 (1985).
6   Hjalmarson, H. P., Vogl, P., Wolford, D. J. & Dow, J. D. Theory of substitutional deep traps in covalent semiconductors. *Physical Review Letters* **44**, 810 (1980).
7   Walukiewicz, W. Amphoteric native defects in semiconductors. *Applied physics letters* **54**, 2094-2096 (1989).
8   Mooney, P. Deep donor levels (DX centers) in III‐V semiconductors. *Journal of Applied Physics* **67**, R1-R26 (1990).
9   Barja, S. *et al.* Identifying substitutional oxygen as a prolific point defect in monolayer transition metal dichalcogenides. *Nature communications* **10**, 1-8 (2019).
10  Cui, X. *et al.* Multi-terminal transport measurements of MoS 2 using a van der Waals heterostructure device platform. *Nature nanotechnology* **10**, 534 (2015).
11  Tian, X. *et al.* Correlating the three-dimensional atomic defects and electronic properties of two-dimensional transition metal dichalcogenides. *Nature Materials*, 1-7 (2020).
12  Amit, I. *et al.* Role of charge traps in the performance of atomically thin transistors. *Advanced Materials* **29**, 1605598 (2017).
13  Yin, L. *et al.* Robust trap effect in transition metal dichalcogenides for advanced multifunctional devices. *Nature communications* **10**, 1-8 (2019).
14  Ramasubramaniam, A. & Naveh, D. Mn-doped monolayer MoS 2: an atomically thin dilute magnetic semiconductor. *Physical Review B* **87**, 195201 (2013).
15  Le, D., Rawal, T. B. & Rahman, T. S. Single-layer MoS2 with sulfur vacancies: structure and catalytic application. *The Journal of Physical Chemistry C* **118**, 5346-5351 (2014).
16  Xie, J. *et al.* Defect‐rich MoS2 ultrathin nanosheets with additional active edge sites for enhanced electrocatalytic hydrogen evolution. *Advanced materials* **25**, 5807-5813 (2013).
17  Vancsó, P. *et al.* The intrinsic defect structure of exfoliated MoS 2 single layers revealed by Scanning Tunneling Microscopy. *Scientific reports* **6**, 29726 (2016).
18  Addou, R., Colombo, L. & Wallace, R. M. Surface defects on natural MoS2. *ACS applied materials & interfaces* **7**, 11921-11929 (2015).
19  Liu, X., Balla, I., Bergeron, H. & Hersam, M. C. Point defects and grain boundaries in rotationally commensurate MoS2 on epitaxial graphene. *The Journal of Physical Chemistry C* **120**, 20798-20805 (2016).
20  Jeong, T. Y. *et al.* Spectroscopic studies of atomic defects and bandgap renormalization in semiconducting monolayer transition metal dichalcogenides. *Nature communications* **10**, 1-10 (2019).
21  Hong, J. *et al.* Exploring atomic defects in molybdenum disulphide monolayers. *Nature communications* **6**, 1-8 (2015).
22  Lang, D. Deep‐level transient spectroscopy: A new method to characterize traps in semiconductors. *Journal of applied physics* **45**, 3023-3032 (1974).
23  McCluskey, M. D. & Haller, E. E. *Dopants and defects in semiconductors*. (CRC press, 2018).
24  Liu, Y. *et al.* Approaching the Schottky–Mott limit in van der Waals metal–semiconductor junctions. *Nature* **557**, 696-700 (2018).

25    Borsuk, J. & Swanson, R. Current transient spectroscopy: A high-sensitivity DLTS system. *IEEE Transactions on electron devices* **27**, 2217-2225 (1980).
26    Blood, P. & Orton, J. W. *The electrical characterization of semiconductors: majority carriers and electron states*. Vol. 2 (Academic press London, 1992).
27    Almbladh, C.-O. & Rees, G. Statistical mechanics of electronic energy levels in semiconductors. *Solid State Communications* **41**, 173-176 (1982).
28    Lagowski, J., Kaminska, M., Parsey Jr, J., Gatos, H. & Lichtensteiger, M. Passivation of the dominant deep level (EL2) in GaAs by hydrogen. *Applied Physics Letters* **41**, 1078-1080 (1982).
29    Vertiatchikh, A., Eastman, L., Schaff, W. & Prunty, T. Effect of surface passivation of AlGaN/GaN heterostructure field-effect transistor. *Electronics Letters* **38**, 388-389 (2002).
30    Yu, P. Y. & Cardona, M. *Fundamentals of semiconductors: physics and materials properties*. (Springer, 1996).
31    Kang, J., Tongay, S., Zhou, J., Li, J. & Wu, J. Band offsets and heterostructures of two-dimensional semiconductors. *Applied Physics Letters* **102**, 012111 (2013).
32    Ci, P. *et al.* Quantifying van der Waals interactions in layered transition metal dichalcogenides from pressure-enhanced valence band splitting. *Nano letters* **17**, 4982-4988 (2017).
33    Claeys, C. & Simoen, E. *Germanium-based technologies: from materials to devices*. (elsevier, 2011).
34    Li, J. *et al.* Nature of Mg impurities in GaN. *Applied physics letters* **69**, 1474-1476 (1996).
35    Lin, J., Dissanayake, A., Brown, G. & Jiang, H. Relaxation of persistent photoconductivity in Al 0.3 Ga 0.7 As. *Physical Review B* **42**, 5855 (1990).
36    McCluskey, M. *et al.* Metastability of oxygen donors in AlGaN. *Physical Review Letters* **80**, 4008 (1998).
37    Chadi, D. & Chang, K.-J. Theory of the atomic and electronic structure of DX centers in GaAs and Al x Ga 1− x As alloys. *Physical review letters* **61**, 873 (1988).
38    Chand, N. *et al.* Comprehensive analysis of Si-doped Al x Ga 1− x As (x= 0 to 1): Theory and experiments. *Physical Review B* **30**, 4481 (1984).
39    Chadi, D. & Chang, K.-J. Energetics of DX-center formation in GaAs and Al x Ga 1− x As alloys. *Physical Review B* **39**, 10063 (1989).
40    Singh, A. & Singh, A. K. Origin of n-type conductivity of monolayer MoS 2. *Physical Review B* **99**, 121201 (2019).
41    Fujisawa, T., Krištofik, J., Yoshino, J. & Kukimoto, H. Metastable behavior of the DX center in Si-doped GaAs. *Japanese journal of applied physics* **27**, L2373 (1988).
42    Tachikawa, M. *et al.* Observation of the persistent photoconductivity due to the DX center in GaAs under hydrostatic pressure. *Japanese journal of applied physics* **24**, L893 (1985).
43    Radisavljevic, B., Radenovic, A., Brivio, J., Giacometti, V. & Kis, A. Single-layer MoS 2 transistors. *Nature nanotechnology* **6**, 147 (2011).
44    Chen, Y. *et al.* Tuning electronic structure of single layer MoS2 through defect and interface engineering. *ACS nano* **12**, 2569-2579 (2018).
45    Liu, G.-B., Xiao, D., Yao, Y., Xu, X. & Yao, W. Electronic structures and theoretical modelling of two-dimensional group-VIB transition metal dichalcogenides. *Chemical Society Reviews* **44**, 2643-2663 (2015).
46    Guo, Y. & Robertson, J. Band engineering in transition metal dichalcogenides: Stacked versus lateral heterostructures. *Applied Physics Letters* **108**, 233104 (2016).
47    Wang, G. *et al.* Spin-orbit engineering in transition metal dichalcogenide alloy monolayers. *Nature communications* **6**, 1-7 (2015).
48    Dabov, K., Foi, A., Katkovnik, V. & Egiazarian, K. Image denoising by sparse 3-D transform-domain collaborative filtering. *IEEE Transactions on image processing* **16**, 2080-2095 (2007).
49    Kresse, G. & Furthmüller, J. Efficient iterative schemes for ab initio total-energy calculations using a plane-wave basis set. *Physical Review B* **54**, 11169 (1996).
50    Blöchl, P. E. Projector augmented-wave method. *Physical review B* **50**, 17953 (1994).
51    Perdew, J. P., Burke, K. & Ernzerhof, M. Generalized gradient approximation made simple. *Physical review letters* **77**, 3865 (1996).

52  Grimme, S. Semiempirical GGA‐type density functional constructed with a long‐range dispersion correction. *Journal of computational chemistry* **27**, 1787-1799 (2006).
53  Wei, S.-H. Overcoming the doping bottleneck in semiconductors. *Computational Materials Science* **30**, 337-348 (2004).

**Acknowledgements**

This work was supported by the Electronic Materials Program funded by the Director, Office of Science, Office of Basic Energy Sciences, Materials Sciences and Engineering Division, of the U.S. Department of Energy under Contract No. DE-AC02-05CH11231. The device fabrication was partly supported by the Center for Energy Efficient Electronics Science (NSF Award No. 0939514). J. M. and X. T. acknowledge the support by the US Department of Energy, Office of Science, Basic Energy Sciences, Division of Materials Sciences and Engineering under award DE-SC0010378 and by an Army Research Office MURI grant on Ab-Initio Solid-State Quantum Materials: Design, Production and Characterization at the Atomic Scale (18057522). We are grateful for Prof. Mary Scott and Dr. Yaqian Zhang for assistance in TEM, and Dr. Muhua Sun for drawing the schematic of the DLTS device.

**Author contributions**

P.C., O. D. and J. W. conceived this project. P. C. fabricated DLTS and PPC devices and completed the measurements, with the assistance from A. S., K. E., S. W., K. T., J. L., Y. C. and O. D.. X. T. and J. M. contributed to atomic-resolution STEM imaging. S. T. grew the bulk $MoS_2$, $WS_2$ and alloys. J. K. performed DFT calculations. P. C., J. K., W. W. and J. W. analyzed the results. All authors discussed and contributed to the preparation of the manuscript.

**Competing interests**

The authors declare no competing interests.

**Additional information**

Supplementary information is available.

# Supplementary Information

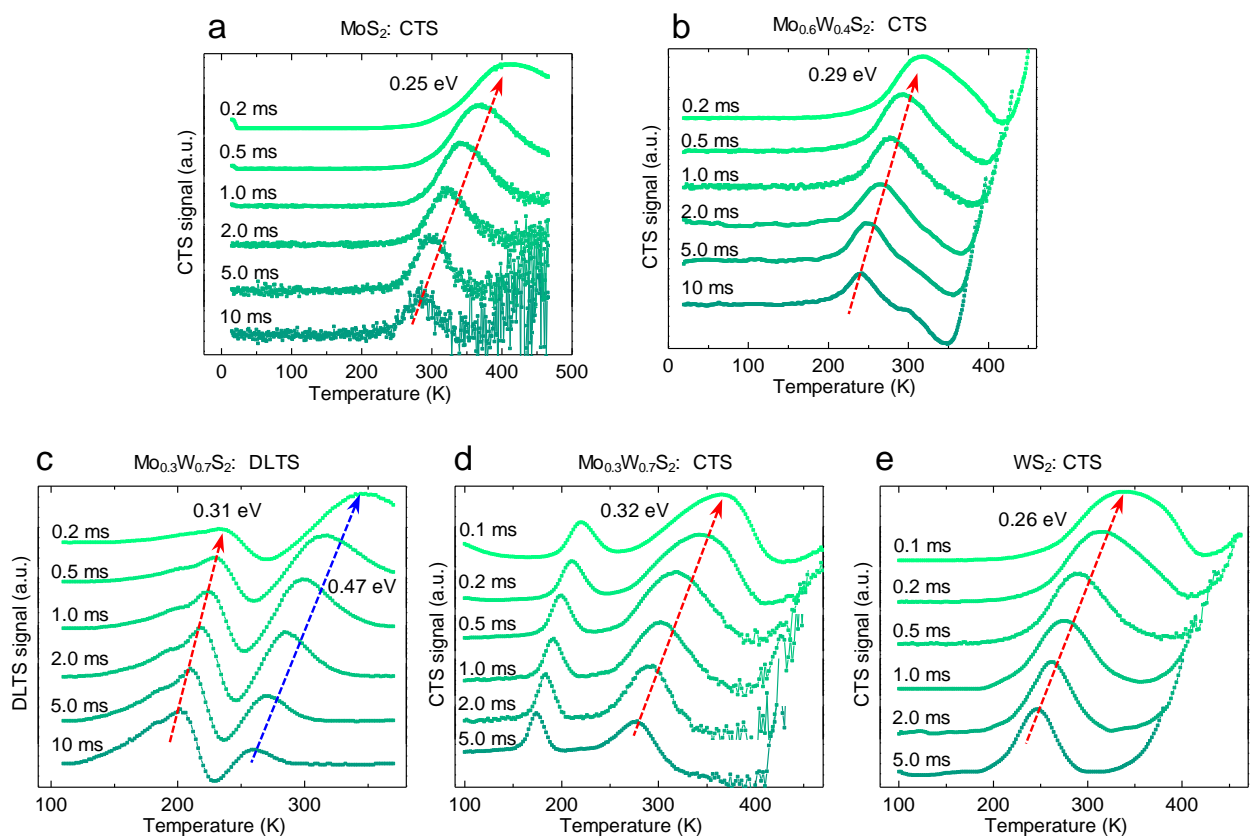

**Supplementary Figure 1: DLTS and CTS data for $Mo_{1-x}W_xS_2$ crystals.** The CTS spectra shows only the $V_S$ feature but with asymmetrical profile, in particular for curves with the rate window below 1.0 ms. The asymmetric shoulder at higher temperatures in the CTS spectra suggests a deeper energy level too weak to be analyzed, and is possibly caused by the DX center. The CTS spectrum of $Mo_{0.3}W_{0.7}S_2$ in Supplementary Fig. 1d shows an additional feature at lower temperatures than that with the activation energy of 0.32 eV. Its origin is currently unknown.

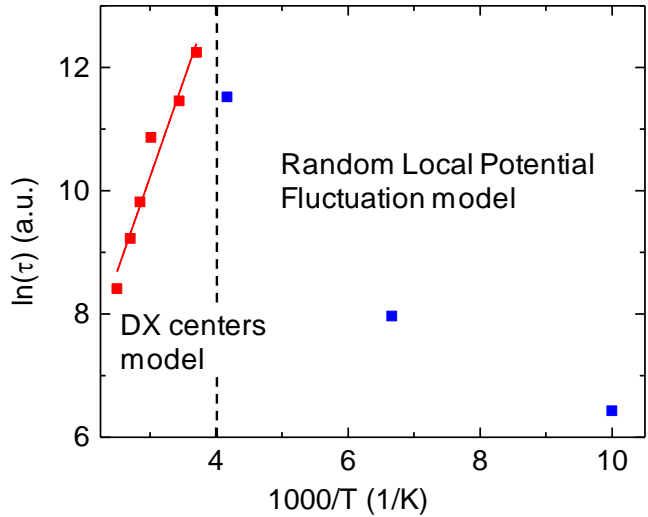

**Supplementary Figure 2: Arrhenius plot of PPC time constant for $MoS_2$ in the broad temperature range.** The DX center model explains the expeditious decay of PPC in the high-temperature regime (T > 270 K). When T < 240 K, the time constant of $MoS_2$ instead drops with decrease in temperature, which is explained by random local potential fluctuation (RLPF), a model that was proposed in previous reports[1,2]. In the RLPF model, random low-potential sites in the conduction band are spatially separated from those in the valence band, so photo-generated electrons (holes) tend to be trapped by these localized sites in the conduction (valence) band, which results in a long carrier lifetime and the PPC effect. These local potential fluctuation in the vdW materials could arise from disordered, charged native defects[2], or randomly distributed trapped charges on the $SiO_2$ substrate[3]. In details, at low temperatures (below ~100 K), photo-excited carriers are confined into these local sites and only contribute weakly to the current flow by hopping transport, hence leading to negligible PPC. As the temperature increases, more electrons gain sufficient kinetic energy to transfer from the localized states to delocalized states, forming a percolation network and thus contributing more to the conductivity, so the PPC effect becomes stronger and decays more slowly [1,4]. Note that when the thermal energy is sufficiently high, it excites all localized electrons from the local potential sites, consequently the PPC effect tends to saturate, and its time constant becomes fixed or only weakly depend on temperature[1].

Indeed, the PPC effect can arise from more than one mechanism. For example, in ZnCdSe[4], in the temperature range from 70 K to 220 K, the time constant of PPC rises with temperature because of the RLPF effect; but when T > 220 K, the time constant shows an opposite temperature dependence: the PPC decays faster as temperature grows. Both DX center and RLPF can cause the PPC effect but with distinct temperature dependencies. The energy barrier, $E_c$, of DX centers prevents photogenerated electrons from transferring to the localized DX centers, hence high temperature expedites the decay of the PPC. In contrast, in the RLPF mechanism, the local potential sites in the conduction band trap electrons that are frozen-out at low temperatures, thus, unlike DX centers, RLPF causes a faster decay of the PPC effect at lower temperatures.

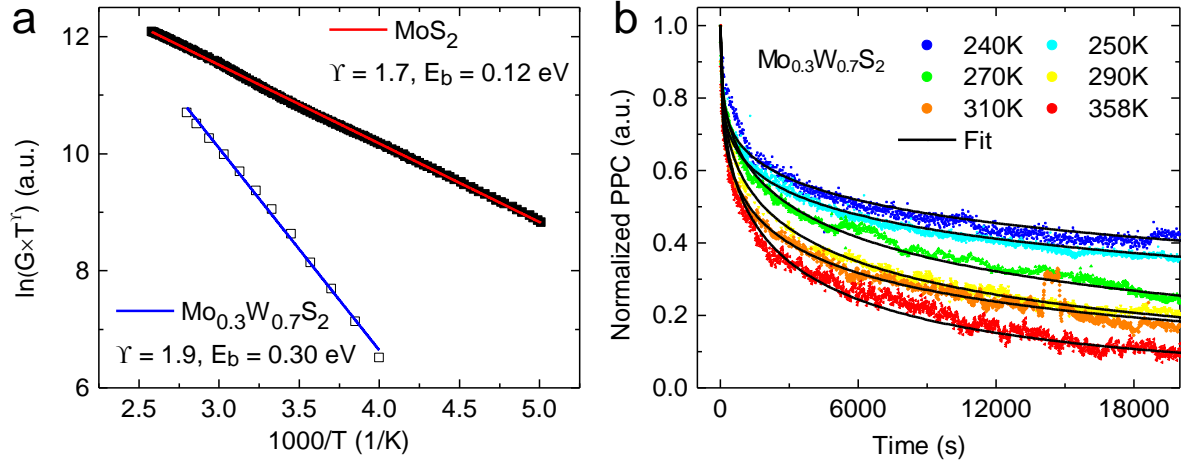

**Supplementary Figure 3**: **Temperature-dependent conductivity and photoconductivity. a**, Arrhenius plots of dark electron density ($n \propto G \times T^\gamma$) for $MoS_2$ and $Mo_{0.3}W_{0.7}S_2$ to determine binding energy $E_b$. **b**, Transient normalized-PPC curves at various temperatures for $Mo_{0.3}W_{0.7}S_2$, where the time constant is extracted by fitting with the stretched-exponential equation and shown in Fig. 3b.

We extract the thermal activation energy ($E_b$) for deep levels in $MoS_2$ and $Mo_{0.3}W_{0.7}S_2$ from the Arrhenius plot of the dark conductance versus inverse temperature as in Supplementary Fig. 3a. The carrier density ($n$) depends exponentially on temperature, $n \sim \exp(-E_b/k_B T)$, considering the Boltzmann distribution and the "full-slope" regime in the freeze-out curves of semiconductors, where only a small portion of the deep levels are ionized [5,6]. On the other hand, conductivity can be expressed by the Drude model as $\sigma = n e \mu$, where $\mu$ is the mobility following a temperature dependence of $\mu \sim T^{-\gamma}$ above ~ 200 K as reported in previous studies[7-9]. Combining these equations, $n$ is related to the conductance G and expressed as,

$$G \times T^\gamma \propto n \propto \exp\left(-\frac{E_b}{k_B T}\right). \quad (1)$$

Arrhenius plot of Supplementary Eq. (1) yields an activation energy of 0.12 eV for $MoS_2$ by using an exponent of $\gamma = 1.7$ as reported in literature[8]. We also found that the obtained value of $E_b$ is insensitive to the value of $\gamma$ used in the fitting, and only changes from 0.09 to 0.13 eV when $\gamma$ is changed from 0.5 to 2.5. Similarly, we extract the activation energy of 0.30eV for $Mo_{0.3}W_{0.7}S_2$ in Supplementary Fig. 3a, where $\gamma$ uses the interpolated value of 1.9 following the known values of 1.7 for $MoS_2$ and 2.0 for $WS_2$[8,9].

These $E_b$ values (0.12 eV for $MoS_2$ and 0.3 eV for $Mo_{0.3}W_{0.7}S_2$) are shallower than the energy level $E_i$ for sulfur vacancy ($V_S$) measured from $E_{CB}$. Therefore, they must originate from a defect level other than the $V_S$. As the only other deep level identified from DLTS is the DX centers, we assign the extracted $E_b$ to the energy distance between $E_{CB}$ and the $E_{DX}$. That is, $E_b = E_{CB} - E_{DX} = E_e - E_c$.

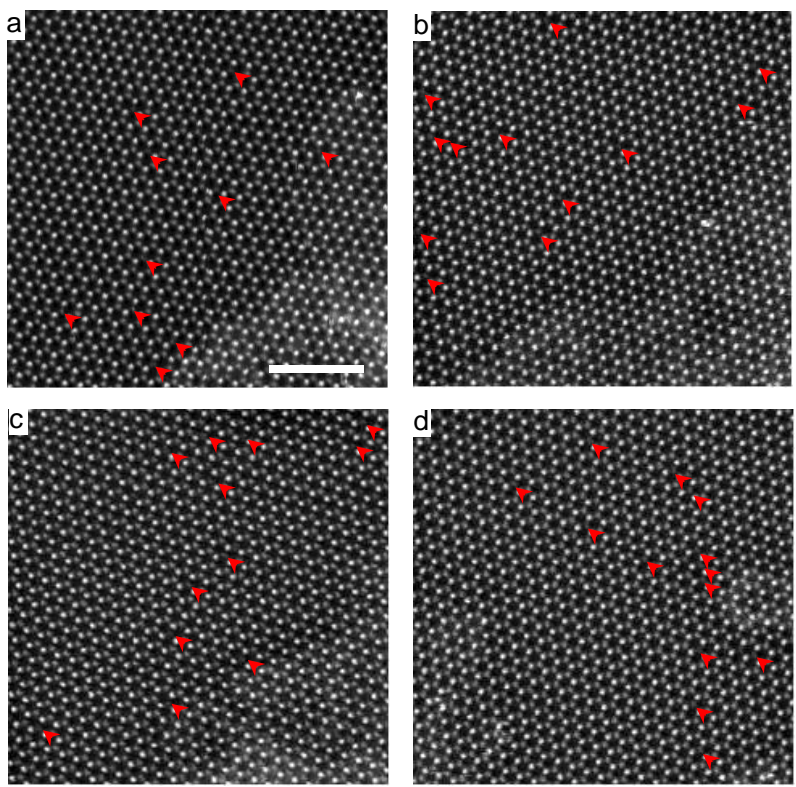

**Supplementary Figure 4: Imaging V$_S$ in monolayer MoS$_2$ with STEM.** Scale bar, 2nm. V$_S$ density is ~ 0.2 nm$^{-2}$ in our exfoliated monolayer MoS$_2$, corresponding to ~ 3×10$^{20}$ cm$^{-3}$ in multilayers. Subtracting the S vacancies induced by the electron beam yields the native density of > 1×10$^{20}$ cm$^{-3}$ (see methods in the main text), in agreement with literature[10,11].

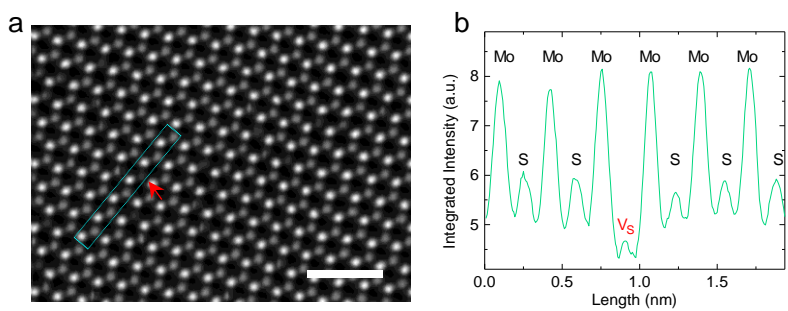

**Supplementary Figure 5: Intensity profiles of V$_S$ in b, corresponding to the boxed region in a.** Scale bar, 1 nm.

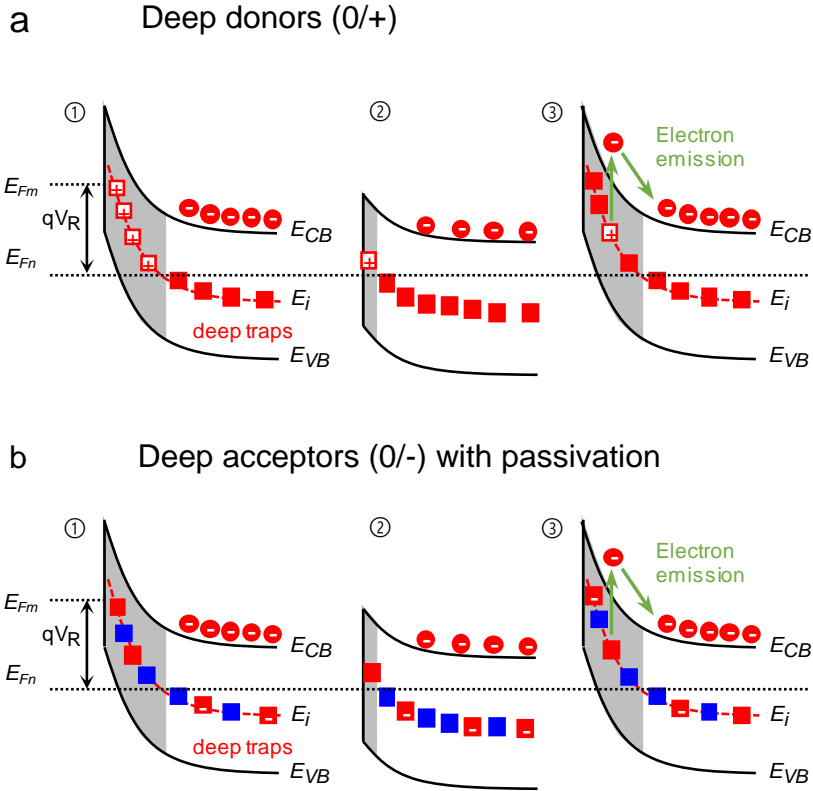

**Supplementary Figure 6: Band bending of a n-type Schottky junction in response to the biased voltage for deep donors in a, and partly passivated deep acceptors in b.** Blue blocks represent passivated (hence inactive and always neutral) deep acceptors. Stage ① - ③ correspond to those in Fig. 1c.

The concentration of $V_S$ determined from STEM in Supplementary Fig. 4 is about $1\times10^{20}$ cm$^{-3}$ in MoS$_2$, which is much higher than the free electron density of not intentionally doped MoS$_2$ on the order of ~ $10^{18}$ cm$^{-3}$ as reported in literature[7,12]. This could be attributed to either compensation or passivation of the deep levels, as widely observed and reported in many traditional semiconductors[13-15]. Supplementary Figure 6a presents the band bending of deep donors with full occupancy in the ground state, akin to the schematic in the main text. However, deep traps in semiconductors may be passivated, and hence de-activated at equilibrium, as shown in the case of deep acceptors in Supplementary Fig. 6b, leading to only a small portion of traps in the depletion zone being active and able to emit electrons under the reverse bias. The mechanism of passivation of the deep levels is currently unknown in MoS$_2$, and is beyond the scope of this study, but the small capture cross section of $V_S$ deep acceptors and their weak attraction to free electrons may play a role.

**Supplementary Note 1: Calculation of deep level capture cross section from DLTS.**

Rewriting Eq. (1) yields $ln\left(\frac{T^2}{e_n}\right) = ln\left(\frac{1}{K \cdot \sigma_n}\right) + \left(\frac{E_{CB}-E_i}{k}\right)\frac{1000}{T}$, where the extrapolation in the Arrhenius plot (Fig. 2b) allows extraction of the capture cross section, $\sigma_n$. The constant $K$ is expressed as[5,16]

$$K = 2\left(\frac{2\pi m_e^* k}{h^2}\right)^{3/2}\left(\frac{3k}{m_{tc}^*}\right)^{1/2} = 3.26 \times 10^{21}\left[\frac{1}{cm^2 K^2 s}\right] \times \left(\frac{m_e^{*3}}{m_{tc}^*}\right)^{1/2}, \quad (2)$$

where $m_{tc}^*$ is the normalized thermal velocity effective mass, and $m_e^*$ is the normalized density of states mass. The latter mass has been determined to be 0.50 (normalized to the free electron mass) as reported by previous studies[17]. The former mass is expressed as[18]

$$m_{tc}^* = \frac{4m_l}{\left[1+\sqrt{m_l/m_t}\sin^{-1}(\delta)/\delta^2\right]^2}, \quad (3)$$

where $\delta = \sqrt{(m_l - m_t)/m_l}$, and $m_l$ and $m_t$ are the longitudinal and transverse effective masses in the ellipsoidal energy surface[18]. Our DFT calculation determines $m_l$ and $m_t$ to be 0.62 and 0.55, respectively, hence giving $m_{tc}^* = 0.57$. Finally, the capture cross section of $V_S$ is calculated to be ~3.6×10$^{-18}$ cm$^2$ in MoS$_2$.

**Supplementary Note 2: Impact of our results: prediction of deep levels of anion impurities.**

The knowledge of $V_S$ attained in this study can be used to understand and predict energy levels of anion-substitutional impurities such as oxygen in $MoS_2$ or $WS_2$. In order to explain this prediction, we start with discussing the bonding / antibonding model for a di-atomic system with the secular equation[19,20]

$$\begin{vmatrix} E - E_{A0} & V \\ V & E - E_{B0} \end{vmatrix} = 0, \qquad (4)$$

where $E_{A0}$ and $E_{B0}$ (lower than $E_{A0}$) are the atomic levels, and $V$ is the interaction between $E_{A0}$ and $E_{B0}$ arising from the wavefunction overlap. Solving this equation yields two eigenvalues, corresponding to the molecular bonding and antibonding energy levels:

$$E_A = \frac{E_{A0} + E_{B0}}{2} + \frac{1}{2}\sqrt{(E_{A0} - E_{B0})^2 + 4V^2}, \qquad (5)$$

and

$$E_B = \frac{E_{A0} + E_{B0}}{2} - \frac{1}{2}\sqrt{(E_{A0} - E_{B0})^2 + 4V^2}. \qquad (6)$$

Then rewriting of Supplementary Eq. (4) gives

$$\Delta' = \frac{1}{2}\left(\sqrt{\Delta_0^2 + 4V^2} - \Delta_0\right), \qquad (7)$$

where $\Delta_0 = E_{A0} - E_{B0} > 0$ represents the energy difference between the initial atomic levels, and $\Delta' = E_A - E_{A0} > 0$ is the difference in energy between the atomic level and its originated molecular orbital (Supplementary Fig. 7). In order to determine the evolution of Supplementary Eq. (7) with the change in $\Delta_0$, the first-order differentiation is calculated as

$$\frac{d\Delta'}{d\Delta_0} = \frac{1}{2}\left(\frac{\Delta_0}{\sqrt{\Delta_0^2 + 4V^2}} - 1\right) < 0, \qquad (8)$$

indicating a monotonically decreasing function of Supplementary Eq. (7). This suggests that increase in the difference of the initial atomic levels will reduce the splitting between the atomic and molecular levels ($\Delta'$, see Supplementary Fig. 7).

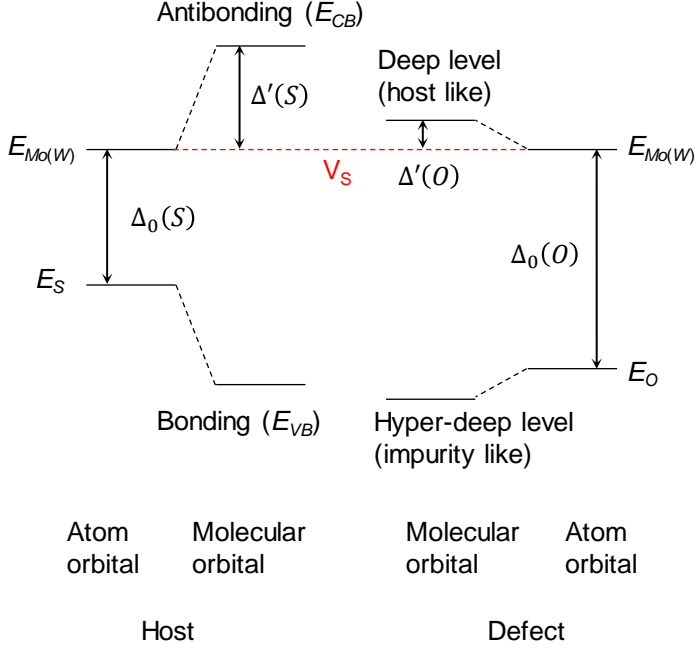

**Supplementary Figure 7: Bonding / antibonding model explaining the formation of deep levels in the bandgap of MoWS$_2$.** The left-hand side is schematic of the atomic levels of Mo(W) and S atoms and the resultant antibonding and bonding states in MoS$_2$ (WS$_2$). The right-hand side shows atomic level of the Mo(W) atom and substitutional O atom (defect) and the resultant deep and hyper-deep levels in MoS$_2$ (WS$_2$) with O defects.

Next, we discuss the formation of the energy level induced by more electronegative, anion substitutional impurities in MoS$_2$ or WS$_2$ in Supplementary Fig. 7. The left-hand side presents the creation of conduction and valence bands in MoS$_2$ following the simplest possible bonding / antibonding model. We note that, according to the origins of conduction band maximum (CBM) and the valence band minimum (VBM) in MoS$_2$ or WS$_2$[21], the antibonding state ($E_{CB}$) in Supplementary Fig. 7 can be defined as the CBM, while the bonding state ($E_{VB}$) is deeper than the VBM, so the difference between $E_{CB}$ and $E_{VB}$ is not equal to the bandgap. The energy difference between the atomic level of Mo(W) atom and the formed conduction band of Mo(W)S$_2$, $\Delta'(S)$, can be expressed by Supplementary Eq. (7). As calculated by DFT in Fig. 2d, the wavefunction of V$_S$ is composed mainly of orbitals of Mo(W) atoms, hence it is reasonable to assume that the position of V$_S$ level lies very close to the atomic level of Mo(W), $E_{Mo(W)}$, in Supplementary Fig. 7. Considering anion impurities such as oxygen substituting S in MoWS$_2$, the interaction between the O atom and its neighboring Mo(W) atoms forms two molecular levels, a deep level and a so-called hyper-deep level[22]. The latter is below the valence band and electrically inactive; in contrast, the former lies inside the bandgap and its wavefunctions is dominated by that of Mo(W), so it is called host-like defect level as shown in the right-hand side of Supplementary Fig. 7, akin to the nitrogen defect in GaP[22,23].

The low-lying oxygen atomic level with respect to the vacuum level means a more significant difference in the original energies ($\Delta_0(O) = E_{Mo} - E_O$) than that in host materials ($\Delta_0(S) = E_{Mo} - E_S$), resulting in the smaller splitting $\Delta'(O)$ in Supplementary Fig. 7, following the Supplementary Eq. (7) and (8). Due to the high electronegativity of O atom, the Mo(W)-O can form a more ionic bond with weaker wavefunction overlap and hence a smaller value of $V$ (Supplementary Eq. (7)), compared to the more covalent Mo(W)-S bond. In summary, it is reasonable to predict that anion impurities would create deep levels with similar energies as the $V_S$, about 0.3 eV below the CBM, in $Mo_{1-x}W_xS_2$ of all compositions.

Finally, the analysis above is not limited to Mo(W) disulfides; all other transition metal chalcogenides may be similarly discussed in the context of native defect energies once the anion vacancy level is measured.

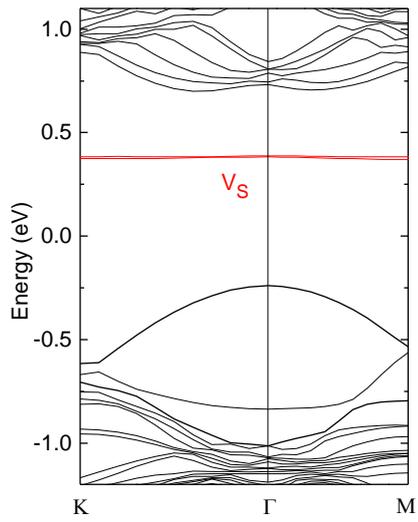

**Supplementary Figure 8: Band structure of multilayer MoS$_2$ with sulfur vacancies by DFT calculations, where V$_S$ indicates the energy level of sulfur vacancies.**

**Supplementary Note 3: Characterization of vdW crystals Field-effect transistors.**

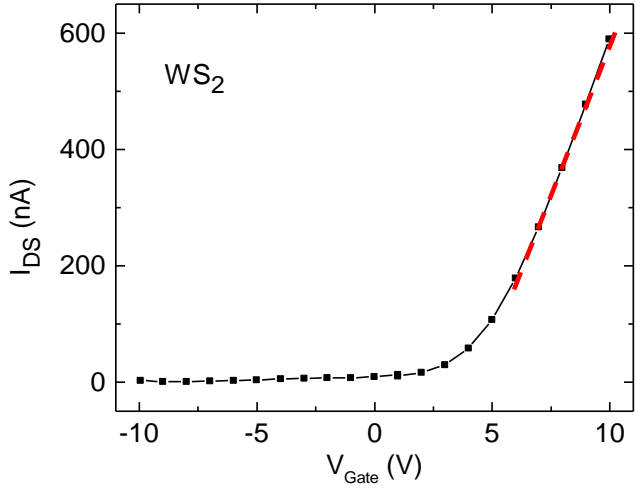

**Supplementary Figure 9: Source-drain current ($I_{DS}$) in response to the gate voltage ($V_{Gate}$) of field-effect transistors (FET) made of $WS_2$ at room temperature.**

It is known that undoped $MoS_2$ has a native electron density over orders of magnitude higher than in $WS_2$[24,25]. Here we explain it using the chemical trend of DX centers in these materials.

According to the data in Fig. 5b, we can extract the low-field field-effect mobility to be ~ 16 cm$^2$/(V·S) for $MoS_2$. Thus, the free carrier density ($n = 1/(e \cdot \mu \cdot \rho)$) of $MoS_2$ is calculated to be ~ $4\times10^{17}$ cm$^{-3}$, consistent with results in previous studies[12,24]. Similarly, the free electron concentration of $WS_2$ is determined to be $7\times10^{13}$ cm$^{-3}$ by the FET results in Supplementary Fig. 9, also in good agreement with literature[25].

Next, we discuss the effect of DX centers on the free carrier density in $MoS_2$ and $WS_2$. We assume the native donor density is on the same level in these two materials, but they are compensated to different extents by the DX centers as deep traps, because of their different energy depths in the bandgap of the hosts. Supplementary Equation 1 is then used to estimate the carrier density ratio of $MoS_2$ to $WS_2$

$$\frac{n(MoS_2)}{n(WS_2)} = exp\left(\frac{E_b(WS_2)-E_b(MoS_2)}{k_BT}\right), \quad (9)$$

where $E_b(MoS_2) = 0.12$ eV is the energy depth of DX centers in $MoS_2$, $E_b(WS_2)$ can be found by extrapolation to be ~ 0.38 eV based on the value of $MoS_2$ and $Mo_{0.3}W_{0.7}S_2$ in Fig.4c, and $k_BT = 26$ meV at room temperature. Therefore, we obtain free carrier density ratio ($n(MoS_2)/n(WS_2)$) to be ~ $2\times10^4$ due to the charge compensation by DX centers. This is on the same order of magnitude as the value ($5\times10^3$) determined from FET measurements.

# Supplementary Note 4: Stray capacitance analysis and capacitance-voltage characterization.

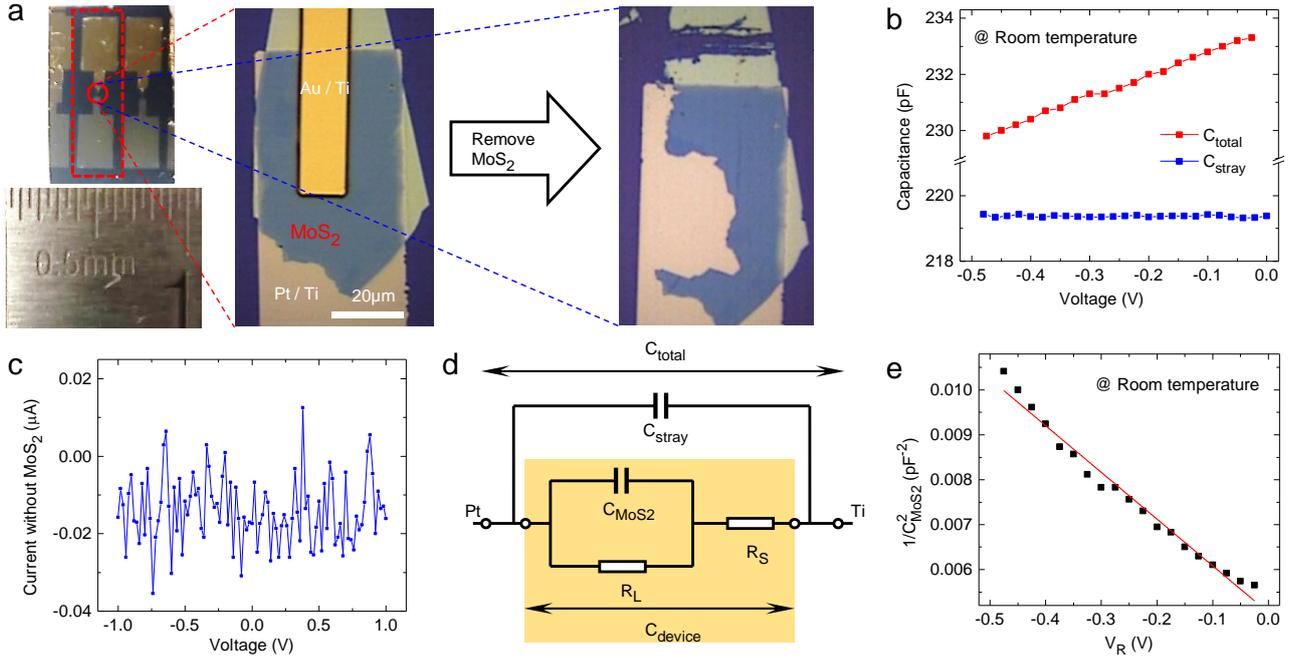

**Supplementary Figure 10: Stray capacitance analysis and capacitance-voltage characterization. a.** Optical image of an empty DLTS device for determination of parasitic capacitance by removing the flake of vdW material with a needle. The red circle represents the stacked parallel capacitor region with / without the flake of vdW material. **b**. Total capacitance ($C_{total}$, with flake) and stray capacitance ($C_{stray}$, removing flake) as a function of reverse bias at room temperature for the device in Supplementary Fig. 10a. **c**. Leakage current of the empty device in response to the bias voltage. **d.** Equivalent circuit of the device in Fig. 1a and Supplementary Fig. 10a. The yellow region represents the Schottky barrier circuit, showing a depletion capacitance of $MoS_2$ ($C_{MoS2}$) with a parallel leakage resistance ($R_L$) and series resistance of the non-depleted region ($R_S$). **e**. $1/C^2_{MoS2}$ vs. reverse voltage to characterize the dopant concentration for the device in Supplementary Fig. 10a at room temperature.

In order to fit the DLTS instrument, the electrodes of devices were designed to be on the size of several millimeters (Supplementary Fig. 10a), but the sample area is about tens of micrometers (Fig. 1a), so the large electrodes inevitably introduce parasitic coplanar capacitance, called stray capacitance, (Supplementary Fig. 10 b and d), whose magnitude is difficult to theoretically estimate due to the irregular geometry. This stray capacitance, even with a large value, is understandably insensitive to the external bias voltage, thereby not affecting the DLTS signal, because the DLTS records the differential capacitance within a rate window under reverse bias (Fig. 1c). Meanwhile, the weak, constant leakage current (~ 0.02 μA, Supplementary Fig. 10c) in the empty device (Supplementary Fig. 10a) ensures the accuracy and reliability of the CTS measurements in Supplementary Fig. 1.

We can determine the $MoS_2$ capacitance by measuring the difference between total capacitance ($C_{total}$, with sample) and stray capacitance ($C_{stray}$, removing sample) in Supplementary Fig. 10 a and b, so the curve of $1/C^2_{MoS2}$ vs. $V_R$ in Supplementary Fig. 10e allows us to obtain the built-in potential ($\Phi_{bi}$) of ~ 0.5 V for MoS$_2$/Pt Schottky diode and the dopant concentration of ~ 3×10$^{18}$ cm$^{-3}$ in nominally not intentionally doped MoS$_2$ ($N_d$) at room temperature, via its intercept and the slope according to[26]

$$\frac{1}{C_{MoS2}^2} = \frac{2(|\Phi_{bi}|+|V_R|)}{qN_d\varepsilon_r\varepsilon_0 A^2}, \quad (10)$$

where $\varepsilon_r = 11$ is the dielectric constant of multilayer MoS$_2$[27], $V_R$ is the reverse biased voltage (Fig. 1c), $A$ is the area of the depletion zone, and C$_{MoS2}$ is close to C$_{device}$ (= C$_{total}$ - C$_{stray}$). These two extracted parameters for MoS$_2$ are consistent with previous results in literature[7,12,28]. As a result, the depletion width of the Schottky junction can be expressed as[26]

$$W = \sqrt{\frac{2\varepsilon_r\varepsilon_0(|\Phi_{bi}|+|V_R|)}{qN_d}}, \quad (11)$$

and estimated to be ~ 22 nm under the reverse bias of 0.5 V at room temperature. Note that the dopant concentration determined here is significantly higher than the free electron density measured by FET in Supplementary Fig. 9, which is attributed to the trapping of free electrons by DX centers.

The linear 1/C$^2$ *vs.* $V_R$ curve indicates a roughly uniform distribution of dopants and nearly step junction profile of space charge density near the surface of the MoS$_2$/Pt Schottky diode[16,29], hence confirming the feasibility of Supplementary Eq. (11) to extract the depletion width.

Although it is reasonable to assume a nearly step junction profile for the space charge, in reality the free carrier density varies exponentially within the depletion zone, so a Debye screening length (or Debye tail, Debye incursion) is defined to express the abruptness of the space charge distribution near the edge of the depletion zone, which can be written as[16]

$$L_D = \sqrt{\frac{\varepsilon_r\varepsilon_0 kT}{q^2 N_d}}. \quad (12)$$

In our case, the high dopant concentration ($N_d$ ~ 3×10$^{18}$ cm$^{-3}$) yields a Debye length of ~ 2 nm at room temperature, which is on the same order with that in heavily doped silicon[16]. The depletion width (~22 nm) is more than ten times greater than this Debye length, which in turn justifies the sharp, nearly step - function profile of space charge[16].

In the above description, we assume that C$_{MoS2}$ is almost equal to C$_{device}$ by omitting the effect of the leakage resistance (R$_L$) and the series resistance (R$_S$). When subtracting the parallel stray capacitance, the measured capacitance, C$_{device}$ in the circuit within the yellow shadow (Supplementary Fig. 10d), is related to C$_{MoS2}$ by[16]

$$\frac{C_{MoS2}}{C_{device}} = \left(1 + \frac{R_S}{R_L}\right)^2 + \left(\frac{R_S}{1/\omega C_{MoS2}}\right)^2, \quad (13)$$

where $\omega$ is the frequency of a.c. voltage during the capacitance measurement and is 1 MHz in our case. Accurate test of the depletion capacitance and hence the depletion width requires that R$_S$ << R$_L$ and R$_S$ << 1/$\omega$C$_{MoS2}$, such that the capacitive impedance, C$_{MoS2}$, dominates the circuit element[16,29,30]. The leakage resistance (R$_L$) and series resistance (R$_S$) can be approximately estimated from the reverse and forward bias current of the Schottky junction to be 80 kΩ and 2.5 kΩ under the reverse bias of 0.2V at 320 K (Fig. 1e), meeting the requirement of R$_S$ << R$_L$. Given that C$_{MoS2}$ ≈ C$_{total}$ − C$_{stray}$, 1/$\omega$C$_{MoS2}$ = 77 kΩ >> R$_S$ at 320 K. Therefore, Supplementary Equation (13) gives C$_{MoS2}$ ~ C$_{device}$, which justifies the reliability of the capacitance measurements. We note that the large leakage current under reverse bias, called 'soft' reverse characteristics, may be attributed to the tunneling effect or the lowering of Schottky barrier height by image forces, as commonly reported in the Schottky junctions formed by low dimensional materials[31-33].

**Supplementary Note 5: Thermodynamic interpretation of Arrhenius plots in DLTS.**

The defect energy level in semiconductors is defined as the change of chemical potential due to the formation of a pair of charged carrier and ionized defect[34,35]. The chemical potential thermodynamically means the variation of Gibbs free energy during the capture or emission of an electron at constant pressure and temperature. Thus, based on these definitions, the Arrhenius equation of the thermal emission rate in Eq. (1) can be rewritten as[35]

$$\frac{e_n}{T^2} = K\sigma_n \exp\left(-\frac{\Delta G(T)}{k_B T}\right), \qquad (14)$$

where $\Delta G(T) = |E_{CB} - E_i|$ and is the activation energy for electron emission from the deep state to the conduction band edge. At the same time, the Gibbs free energy is defined by the thermodynamic identity as $\Delta G(T) = \Delta H - T\Delta S$, where $\Delta H$ and $\Delta S$ represent the changes in enthalpy and entropy, respectively. Therefore, Supplementary Equation (14) becomes[16]

$$\frac{e_n}{T^2} = K\left[\exp\left(\frac{\Delta S}{k_B}\right)\sigma_n\right]\exp\left(-\frac{\Delta H}{k_B T}\right), \qquad (15)$$

and hence the slope of the Arrhenius plot via Eq. (1) yields an average of enthalpy change over the temperature range of this plot, considering the generally weak temperature dependence of $\Delta H$[16]. The difference between $\Delta G$ and $\Delta H$ mainly arises from the lattice vibrational contribution to $\Delta S$ due to the coupling of occupied deep states to the lattice, and therefore, it is usually negligible when electrons are excited from the traps to conduction band without changing the bonding configuration ($\Delta S \sim 0$)[36]. Thus, in this study, it is reasonable to consider the measured Arrhenius slope from DLTS as the activation energy for $V_S$ states, because our DFT calculations do not observe lattice relaxation or entropy change during the transfer of electrons between the $V_S$ defect and the conduction band edge.

With regard to DX centers, most of previous studies on group III-V semiconductors also neglected the difference between $\Delta G$ and $\Delta H$[37-43], despite the occurrence of lattice relaxation when a DX center switches to the electron-donating state. In this study, we do not consider this difference for DX centers in vdW crystals. On the other hand, in order to obtain the exact activation energy, $\Delta G(T)$, via Supplementary Eq. (14), one needs to measure the values of both $e_n$ and $\sigma_n$ at desired temperatures. The emission rate, $e_n$, can be determined by the DLTS or transient capacitance test, while the capture cross section is usually measured using the diode short-circuiting technique[44,45], which is out of the scope of this study.

The main text and Supplementary Note 1 show the extraction of capture cross section, $\sigma_n$, of $V_S$ deep state via the intercept of the Arrhenius plot in Fig. 2b. However, we note that, based on Supplementary Eq. (15), this intercept more accurately represents the product $\exp\left(\frac{\Delta S}{k_B}\right)\sigma_n$, rather than just $\sigma_n$. Experimentally, one could measure the latter using the diode shorting-circuiting technique[44,45] to eventually determine the prefactor, $\Delta S$, by temperature-dependent Gibbs free energy ($\Delta G = \Delta H - T\Delta S$).

**Supplementary Note 6: Activation energy of DX centers by Arrhenius plot of DLTS.**

Unlike regular deep levels which have no capture/emission barriers, such as the $V_S$ state, for DX centers the Arrhenius plot of the DLTS spectrum extracts the emission barrier $E_e$, which is not the energy of the DX center directly measured from the conduction or valence band edges. This is because, in the case of DX centers, the energy barrier, $E_c$ in the configurational coordinate diagram (CCD, Fig. 3c and 4c), must be overcome in order for an electron to be trapped by defects, hence leading to a strongly temperature-dependent capture cross section[5],

$$\sigma_{n,DX} = \sigma_\infty \exp\left(-\frac{E_c}{k_B T}\right). \qquad (16)$$

Combining Supplementary Eq. (16) and Eq. (1) gives

$$\frac{e_n}{T^2} = K\sigma_\infty \exp\left(-\frac{|E_{CB}-E_i|+E_c}{k_B T}\right), \qquad (17)$$

where $|E_{CB} - E_i| + E_c$ is equal to the emission energy, $E_e$, in the CCD (Fig. 3c) without considering the entropy change, and $E_i$ is $E_{DX}$.

To sum up, the DLTS spectrum measures the activation energy or binding energy ($E_b$) for normal defects such as the $V_S$ states, while for DX centers, DLTS yields the emission energy ($E_e$), the summation of binding energy ($E_b$) and capture barrier ($E_c$).

**Supplementary Note 7: Excluding a surface depletion mechanism for the PPC effect.**

Although a surface depletion model was used to explain PPC effects in some low-dimensional systems[46,47], this model is unlikely to explain our observed PPC in $MoS_2$ and alloys. This is because PPC effects induced by surface depletion, for instance, in Si NWs or $\alpha$-$In_2Se_3$ nanosheets[46,47], originate from self-assembled molecules on the surface or oxygen ions adsorbed from the environment. However, our STEM images (Fig. 1b) confirm the absence of adsorbents or contamination on the surface. Moreover, all the PPC tests were completed in high vacuum (~ $10^{-6}$ torr) after annealing at 400 K in vacuum for at least one day to remove possible adsorbents. In the meantime, such surface depletion mechanism usually induces only a weak PPC with a short decay constant (*e.g.*, ~ seconds at room temperature)[46], in stark contrast to the long PPC decay time we observed (~ $10^5$ s for multilayer $MoS_2$ at room temperature). Therefore, our PPC effect observed in $MoS_2$ and alloys is unlikely to be caused by any surface modification of the samples.

## Supplementary References

1 Jiang, H. & Lin, J. Percolation transition of persistent photoconductivity in II-VI mixed crystals. *Physical review letters* **64**, 2547 (1990).

2 Wu, Y.-C. *et al.* Extrinsic origin of persistent photoconductivity in monolayer MoS 2 field effect transistors. *Scientific reports* **5**, 11472 (2015).

3 Xue, J. *et al.* Scanning tunnelling microscopy and spectroscopy of ultra-flat graphene on hexagonal boron nitride. *Nature materials* **10**, 282-285 (2011).

4 Jiang, H. & Lin, J. Persistent photoconductivity and related critical phenomena in Zn 0.3 Cd 0.7 Se. *Physical Review B* **40**, 10025 (1989).

5 McCluskey, M. D. & Haller, E. E. *Dopants and defects in semiconductors*. (CRC press, 2018).

6 Chand, N. *et al.* Comprehensive analysis of Si-doped Al x Ga 1− x As (x= 0 to 1): Theory and experiments. *Physical Review B* **30**, 4481 (1984).

7 Radisavljevic, B. & Kis, A. Mobility engineering and a metal–insulator transition in monolayer MoS 2. *Nature materials* **12**, 815-820 (2013).

8 Perera, M. M. *et al.* Improved carrier mobility in few-layer MoS2 field-effect transistors with ionic-liquid gating. *ACS nano* **7**, 4449-4458 (2013).

9 Xu, S. *et al.* Universal low-temperature Ohmic contacts for quantum transport in transition metal dichalcogenides. *2D Materials* **3**, 021007 (2016).

10 Hong, J. *et al.* Exploring atomic defects in molybdenum disulphide monolayers. *Nature communications* **6**, 1-8 (2015).

11 Qiu, H. *et al.* Hopping transport through defect-induced localized states in molybdenum disulphide. *Nature communications* **4**, 1-6 (2013).

12 Radisavljevic, B., Radenovic, A., Brivio, J., Giacometti, V. & Kis, A. Single-layer MoS 2 transistors. *Nature nanotechnology* **6**, 147 (2011).

13 Lagowski, J., Kaminska, M., Parsey Jr, J., Gatos, H. & Lichtensteiger, M. Passivation of the dominant deep level (EL2) in GaAs by hydrogen. *Applied Physics Letters* **41**, 1078-1080 (1982).

14 Dautremont‐Smith, W. *et al.* Passivation of deep level defects in molecular beam epitaxial GaAs by hydrogen plasma exposure. *Applied physics letters* **49**, 1098-1100 (1986).

15 Nabity, J. *et al.* Passivation of Si donors and DX centers in AlGaAs by hydrogen plasma exposure. *Applied physics letters* **50**, 921-923 (1987).

16 Blood, P. & Orton, J. W. *The electrical characterization of semiconductors: majority carriers and electron states*. Vol. 2 (Academic press London, 1992).

17 Peelaers, H. & Van de Walle, C. G. Effects of strain on band structure and effective masses in MoS 2. *Physical Review B* **86**, 241401 (2012).

18 Green, M. A. Intrinsic concentration, effective densities of states, and effective mass in silicon. *Journal of Applied Physics* **67**, 2944-2954 (1990).

19 Ci, P. *et al.* Quantifying van der Waals interactions in layered transition metal dichalcogenides from pressure-enhanced valence band splitting. *Nano letters* **17**, 4982-4988 (2017).

20 Burns, G. *Solid State Physics*. (Elsevier Science, 1985).

21 Kang, J., Tongay, S., Zhou, J., Li, J. & Wu, J. Band offsets and heterostructures of two-dimensional semiconductors. *Applied Physics Letters* **102**, 012111 (2013).

22 Hjalmarson, H. P., Vogl, P., Wolford, D. J. & Dow, J. D. Theory of substitutional deep traps in covalent semiconductors. *Physical Review Letters* **44**, 810 (1980).

23 Yu, P. Y. & Cardona, M. *Fundamentals of semiconductors: physics and materials properties*. (Springer, 1996).

24 Kwak, J. Y. *et al.* Electrical characteristics of multilayer MoS2 FET's with MoS2/graphene heterojunction contacts. *Nano letters* **14**, 4511-4516 (2014).

25 Braga, D., Gutiérrez Lezama, I., Berger, H. & Morpurgo, A. F. Quantitative determination of the band gap of WS2 with ambipolar ionic liquid-gated transistors. *Nano letters* **12**, 5218-5223 (2012).

26 Hu, C. *Modern semiconductor devices for integrated circuits*. Vol. 2 (Prentice Hall Upper Saddle River,

|     | New Jersey, 2010). |
| --- | --- |
| 27  | Chen, X. et al. Probing the electron states and metal-insulator transition mechanisms in molybdenum disulphide vertical heterostructures. *Nature communications* **6**, 6088 (2015). |
| 28  | Kim, G.-S. et al. Schottky barrier height engineering for electrical contacts of multilayered MoS2 transistors with reduction of metal-induced gap states. *ACS nano* **12**, 6292-6300 (2018). |
| 29  | Pierret, R. F. *Semiconductor device fundamentals*.  (Pearson Education India, 1996). |
| 30  | Goodman, A. M. Metal—Semiconductor barrier height measurement by the differential capacitance method—One carrier system. *Journal of Applied Physics* **34**, 329-338 (1963). |
| 31  | Kumar, S. et al. Influence of barrier inhomogeneities on transport properties of Pt/MoS2 Schottky barrier junction. *Journal of Alloys and Compounds* **797**, 582-588 (2019). |
| 32  | Rhoderick, E. H. & Rhoderick, E. *Metal-semiconductor contacts*.  (Clarendon Press Oxford, 1978). |
| 33  | Chen, C.-C., Aykol, M., Chang, C.-C., Levi, A. & Cronin, S. B. Graphene-silicon Schottky diodes. *Nano letters* **11**, 1863-1867 (2011). |
| 34  | Van Vechten, J. & Thurmond, C. Entropy of ionization and temperature variation of ionization levels of defects in semiconductors. *Physical Review B* **14**, 3539 (1976). |
| 35  | Thurmond, C. The standard thermodynamic functions for the formation of electrons and holes in Ge, Si, GaAs, and GaP. *Journal of the Electrochemical Society* **122**, 1133 (1975). |
| 36  | Almbladh, C.-O. & Rees, G. Statistical mechanics of electronic energy levels in semiconductors. *Solid State Communications* **41**, 173-176 (1982). |
| 37  | Criado, J., Gomez, A., Munoz, E. & Calleja, E. Deep level transient spectroscopy signature analysis of DX centers in AlGaAs and GaAsP. *Applied physics letters* **49**, 1790-1792 (1986). |
| 38  | Calleja, E., Gomez, A., Munoz, E. & Camara, P. Fine structure of the alloy‐broadened thermal emission spectra from DX centers in GaAlAs. *Applied physics letters* **52**, 1877-1879 (1988). |
| 39  | Kasu, M., Fujita, S. & Sasaki, A. Observation and characterization of deep donor centers (DX centers) in Si‐doped AlAs. *Journal of applied physics* **66**, 3042-3046 (1989). |
| 40  | Calleja, E., Gomez, A. & Munoz, E. Direct evidence of the DX center link to the L‐conduction‐band minimum in GaAlAs. *Applied physics letters* **52**, 383-385 (1988). |
| 41  | Mooney, P., Theis, T. & Wright, S. Effect of local alloy disorder on emission kinetics of deep donors (DX centers) in Al x Ga1− x As of low Al content. *Applied physics letters* **53**, 2546-2548 (1988). |
| 42  | Kumagai, O., Kawai, H., Mori, Y. & Kaneko, K. Chemical trends in the activation energies of DX centers. *Applied physics letters* **45**, 1322-1323 (1984). |
| 43  | Theis, T., Mooney, P. & Wright, S. Electron Localization by a Metastable Donor Level in n− GaAs: A New Mechanism Limiting the Free-Carrier Density. *Physical review letters* **60**, 361 (1988). |
| 44  | Brotherton, S. & Lowther, J. Electron and hole capture at Au and Pt centers in silicon. *Physical Review Letters* **44**, 606 (1980). |
| 45  | Brotherton, S. & Bicknell, J. The electron capture cross section and energy level of the gold acceptor center in silicon. *Journal of Applied Physics* **49**, 667-671 (1978). |
| 46  | Anandan, M. et al. High-responsivity broad-band sensing and photoconduction mechanism in direct-Gap α-In2Se3 nanosheet photodetectors. *Nanotechnology* **31**, 465201 (2020). |
| 47  | Zhao, X., Tu, P., He, J., Zhu, H. & Dan, Y. Cryogenically probing the surface trap states of single nanowires passivated with self-assembled molecular monolayers. *Nanoscale* **10**, 82-86 (2018). |